# RESOLUTION OF DIAGONALS AND MODULI SPACES

Victor GINZBURG
The University of Chicago,
Department of Mathematics
Chicago, IL 60637, USA
E-mail: ginzburg@zaphod.uchicago.edu

### Table of Contents

## 1. Introduction

This paper is a continuation of [BG]. In that paper, for any smooth complex curve $X$ and $n > 1$, we constructed a canonical completion of the configuration space of all ordered $n$-tuples of distinct points of $X$. The completion is called *Resolution of Diagonals*. There is a natural stratification of the resolution of diagonals with the set of strata being parametrized by certain graphs. The homology groups of the strata turn out to be closely related to free Lie algebras. That relation played a crucial role in [BG]. It was used there for describing jets of functions on moduli spaces of principal $G$-bundles, a higher order analogue of the Kodaira-Spencer isomorphism.

In this paper we extend the results of [BG] in various different ways. First, we show (§3) that the construction of [BG] has a natural counterpart for the moduli space of curves of a fixed genus instead of the moduli space of $G$-bundles.

Second, Fulton and MacPherson have recently constructed [FM] a canonical resolution of diagonals (of the configuration space) for a variety of arbi-trary dimension. That enables us to relate, following the pattern of [BG], the Geometry of the moduli space, $\mathcal{M}$, of holomorphic principal $G$-bundles on an arbitrary compact complex manifold $X$ to the combinatorics and topology of the corresponding resolution of diagonals. The only difference with the 1-dimensional case is that, in general, the results make sense only on the part of the moduli space formed by *stable* bundles. The reason is, that such a moduli space should be really viewed as a *stack* (cf. [La]), and that the stack in question is smooth if $\dim X = 1$ and is no longer smooth if $\dim X > 1$. (Similar results hold for moduli spaces of $G$-bundles with flat connection).

Third, we extend the analysis of [BG] to studying sheaves of algebraic differential operators acting on sections of various natural vector bundles on the moduli space $\mathcal{M}$. There are essentially two different types of such bundles. Vector bundles of the first type are associated to a finite collection of points of $X$ and a finite collection of $G$-modules. In that case a complete description of geometric fibers of the corresponding sheaf of differential operators is given in (§7) below. Vector bundles of the second type are determinant line bundles on $\mathcal{M}$ associated to various $G$-modules. We provide a canonical description of sheaves of first and second order differential operators on the determinant bundles. As a by-product, we get a new construction of the flat projective connection on the vector bundle of conformal blocks [AW]. The projective connection has been recently studied from various points of view ([Hi],[BK],[Fa]). The advantage of our present approach is that it does not appeal to any Global geometry of the moduli spaces and produces, we believe, the most explicit formula for the connection.

The present paper contains only announcements of main results. Complete proofs will be published elsewhere. The reader may consult [GK] for a relevant treatment of the theory of operads which is closely related to the subject of this paper.

It is a pleasure to express my gratitude to Vladimir Drinfeld and Maxim Kontsevich for many stimulating and very useful conversations. In particular, the ideas of Kontsevich are at the origin of my understanding of the relationship between homotopy algebra and quantum field theory. I am also especially indebted to Sasha Beilinson for introducing me to the subject of conformal field theory and of moduli spaces. His contribution was quite essential in obtaining the results of §§$9 - 10$.

## 2. Resolution of Diagonals

Let $X$ be a smooth complex manifold of dimension $d$ and $n > 1$ a positive integer. Let $\mathring{X}^n \subset X^n$ be the open part of all $n$-tuples of pairwise distinct points of $X$, and let $D := X^n \setminus \mathring{X}^n$ be the Diagonal subset. Fulton and MacPherson [FM] have constructed a smooth variety $\hat{X}^n$ and a projective morphism $\pi : \hat{X}^n \twoheadrightarrow X^n$, which is an isomorphism over $\mathring{X}^n$ and such that $\hat{D} := \pi^{-1}(D)$ is a normall crossing divisor.

The simplest way to define $\hat{X}^n$ goes as follows [FM, §1]. Extend our notation, and for any finite set $I$ of cardinality $\#I$, let $X^I (\simeq X^{\#I})$ denote the variety of all $X$-valued functions on $I$, and $\mathring{X}^I \subset X^I$ the corresponding open part of $I$-tuples of distinct points. Let $\Delta \subset X^I$ be the *principal diagonal* formed by constant functions, and let $X_\Delta^I$ denote the blow up of $X^I$ along the subvariety $\Delta$. The subset $\mathring{X}^I$ is not affected by the blow up and, therefore, can be viewed as part of $X_\Delta^I$. Hence, for each subset $I \subset \{1, \ldots, n\}$, $\#I \geq 2$, there are natural maps $\mathring{X}^n \twoheadrightarrow \mathring{X}^I \hookrightarrow X_\Delta^I$. Assembled together, the compositions of those maps give rise to an imbedding:

$$\mathring{X}^n \hookrightarrow \prod_{I \subset \{1\ldots n\}, \#I \geq 2} X_\Delta^I \tag{2.1}$$

It was shown in [FM] that the closure of the image of the above imbedding is a smooth variety, to be denoted $\hat{X}^n$. We often identify $\mathring{X}^n$ with its image in $\hat{X}^n$. Observe next that there is, by construction, a natural projection $\hat{X}^n \twoheadrightarrow X_\Delta^n$. Composing it with the blow down morphism yields a proper morphism $\pi : \hat{X}^n \twoheadrightarrow X^n$. Put $\hat{D} = \pi^{-1}(D)$. It turns out that $\hat{D}$ is a normal crossing divisor with smooth irreducible components.

Any finite collection of smooth divisors with normal crossings on a smooth variety $Y$ gives rise to a filtration by closed subvarieties $Y = Y^0 \supset Y^1 \supset \ldots$, where $Y^k$ is defined to be the union of all intersections of $k$-tuples of distinct divisors from the collection. Then, the connected components of the sets $Y^k \setminus Y^{k+1}$, $k = 0, 1, \ldots$ form a stratification of $Y$ by locally-closed subvarieties with smooth closures. Thus, there is a canonical stratification $\hat{X}^n = \coprod_T S_T$ associated to the collection of divisors $\hat{D}_I \subset \hat{X}^n$, $\#I \geq 2$. The combinatorial structure of the stratification depends neither on $X$ nor even on the dimension of $X$. The set of strata is indexed by *n-groves* [BG, §2], i.e., disjoint unions of oriented trees of certain shape whose external outgoing edges are labeled by the elements $1, \ldots, n$. We have (see [BG, Prop. 3.6]):

$$codim S_T = \text{number of vertices of the grove } T.$$

There is an obvious action of the Symmetric group $\Sigma_n$ on $\mathring{X}^n$ by permutation of factors. The construction of the completion $\hat{X}^n$ shows that the $\Sigma_n$-action on $\mathring{X}^n$ can be extended to a $\Sigma_n$-action on $\hat{X}^n$. The action maps strata into strata. Further, there is a natural $\Sigma_n$-action on the set of $n$-groves by permutation of the labels assigned to external edges of the grove. For any $\sigma \in \Sigma_n$ and any $n$-grove $T$, we have $\sigma(S_T) = S_{\sigma(T)}$. Let $\Sigma(T) \subset \Sigma_n$ and $\Sigma_{in}(T)$ denote respectively the isotropy group of and the permutation group of the set of incoming external edges of a grove $T$ (the latter group is *not* a subgroup of $\Sigma_n$). Both groups act on the stratum $S_T$ in a natural way.

Given a smooth algebraic variety $Y$, let $\Omega^j_Y$ denote the sheaf of regular $j$-forms on $Y$. If $\mathring{Y}$ is a Zariski-open part of $Y$ such that $Y \setminus \mathring{Y}$ is a normal crossing divisor let $\Omega^j_{Y,\mathring{Y}}$ denote the sheaf on $Y$ of $j$-forms with logarithmic singularities at the divisor. Let $D$ be an irreducible component of the divisor and $\mathring{D}$ the complement in $D$ to all other components. Then $D \setminus \mathring{D}$ is a normal crossing divisor in $D$ and, for any $i \geq 0$, there are *residue* maps $H^i(\mathring{Y}, \mathbb{C}) \to H^{i-1}(\mathring{D}, \mathbb{C})$ and $Res : H^i(Y, \Omega^j_{Y,\mathring{Y}}) \to H^i(D, \Omega^{j-1}_{D,\mathring{D}})$. The superscript $j$ is usually dropped from the notation if differential forms of top degree are considered.

## 3. Moduli spaces of curves

Let $\mathcal{M}_g$ denote the moduli space of complex curves of a fixed genus $g > 1$. Let $X$ be a point of $\mathcal{M}_g$, a smooth compact curve of genus $g$. The tangent space to $\mathcal{M}_g$ at the point $X$ is given by the Kodaira-Spencer isomorphism:

$$T_X \mathcal{M}_g \simeq H^1(X, \mathcal{T}_X) \tag{3.1}$$

where $\mathcal{T}_X$ denotes the tangent sheaf on $X$. Further, for an integer $n \geq 1$, let $\mathfrak{J}^n_X$ be the vector space of $n$-jets at $X \in \mathcal{M}_g$ of regular functions on $\mathcal{M}_g$ vanishing at $X$. We have $\mathfrak{J}^1_X = (T_X \mathcal{M}_g)^*$. Hence, (3.1) yields, by Serre duality:

$$\mathfrak{J}^1_X \simeq H^0(X, \Omega^{\otimes 2}_X) \tag{3.2}$$

We generalize (3.2) to higher order jet spaces $\mathfrak{J}_X^n$ as follows. For any $n \geq 1$, let $\Omega^{\otimes 2}_{\hat{X}^n, \mathring{X}^n}$ be the tensor square of the sheaf $\Omega_{\hat{X}^n, \mathring{X}^n}$ of top degree forms on the open stratum $\mathring{X}^n \subset \hat{X}^n$ with logarithmic singularities at the divisor $\hat{D}$. For each stratum $S_T \subset \hat{D}$, choose a family of strata $\mathring{X}^n = S_0, S_1, \ldots, S_k = S_T$ such that $\mathrm{codim} S_i = i$ and such that $S_{i+1}$ is an irreducible component of $\bar{S}_i \setminus S_i$, for any $i = 0, 1, \ldots$ . A choice of the family $S_1, \ldots, S_k$ gives rise to the iterated residue map:

$$H^0(\hat{X}^n, \Omega^{\otimes 2}_{\hat{X}^n, \mathring{X}^n}) \xrightarrow{Res} H^0(\bar{S}_1, \Omega^{\otimes 2}_{\bar{S}_1, S_1}) \xrightarrow{Res} \ldots \xrightarrow{Res} H^0(\bar{S}_T, \Omega^{\otimes 2}_{\bar{S}_T, S_T}) \quad (3.3)$$

The resulting composition map $Res_T : H^0(\hat{X}^n, \Omega^{\otimes 2}_{\hat{X}^n, \mathring{X}^n}) \to H^0(\bar{S}_T, \Omega^{\otimes 2}_{\bar{S}_T, S_T})$ does not depend on the choice of a family $S_1, \ldots, S_k$.

Let $H^\spadesuit(\hat{X}^n, \Omega^{\otimes 2}_{\hat{X}^n, \mathring{X}^n}) \subset H^0(\hat{X}^n, \Omega^{\otimes 2}_{\hat{X}^n, \mathring{X}^n})$ denote the space of all $\Sigma_n$-invariant forms $\omega \in H^0(\hat{X}^n, \Omega^{\otimes 2}_{\hat{X}^n, \mathring{X}^n})$ that satisfy the following condition:

*For each stratum $S_T \subset \hat{X}^n$, the form $Res_T \omega$ is a $\Sigma_{in}(T)$-invariant section of $\Omega^{\otimes 2}_{\bar{S}_T, S_T}$.*

Here is the main result of this section.

**Theorem 3.4.** *For any $n \geq 1$, there is a canonical isomorphism of vector spaces:*

$$\mathfrak{J}_X^n \simeq H^\spadesuit(\hat{X}^n, \Omega^{\otimes 2}_{\hat{X}^n, \mathring{X}^n})$$

## 4. Resolution of Diagonals of a linear space

In this section, we fix an integer $d \geq 1$ and let $\mathbb{A}$ denote a $d+1$-dimensional affine linear space without preferred origin, a principal homogeneous space of the additive group $\mathbb{C}^{d+1}$. Let $\mathrm{Aff} \simeq \mathbb{C}^* \ltimes \mathbb{C}^{d+1}$ denote the group of affine transformations of $\mathbb{A}$ with scalar linear part. Further, let $n$ be another positive integer and $\mathbb{A}^n$ the $n$-th cartesian power of $\mathbb{A}$. The group Aff acts diagonally on $\mathbb{A}^n$ and the principal diagonal $\Delta \subset \mathbb{A}^n$ is an Aff-stable subvariety. The Aff-action on $\mathbb{A}^n \setminus \Delta$ is free and the corresponding orbit space, $\mathbb{P}^{d,n}$, is isomorphic to a projective space:

$$\mathbb{P}^{d,n} := (\mathbb{A}^n \setminus \Delta)/\mathrm{Aff} \simeq \mathbb{P}^{(d+1) \cdot n - d - 1}$$

Let $\mathring{\mathbb{P}}^{d,n}$ denote the image of $\mathring{\mathbb{A}}^n$ under the quotient map $\mathbb{A}^n \setminus \Delta \twoheadrightarrow \mathbb{P}^{d,n}$.

The construction of §2 applied to the variety $X = \mathbb{A}$ produces the resolution of diagonals $\hat{\mathbb{A}}^n \twoheadrightarrow \mathbb{A}^n$. Let $\hat{\Delta}$ be the inverse image of $\Delta$ in $\hat{\mathbb{A}}^n$, a component of the divisor $\hat{D}$. The diagonal Aff-action on the open part $\mathring{\hat{\mathbb{A}}}^n$ can be extended, by continuity, to an Aff-action on the whole of $\hat{\mathbb{A}}^n$ so that it commutes with the projection to $\mathbb{A}^n$. It follows, that the restriction of the Aff-action to $\hat{\mathbb{A}}^n \setminus \hat{\Delta}$ is free so that the corresponding orbit space has a natural structure of a smooth projective variety, $\hat{\mathbb{P}}^{d,n}$. Furthermore, there is a natural morphism

$$\hat{\mathbb{P}}^{d,n} := (\hat{\mathbb{A}}^n \setminus \hat{\Delta})/\text{Aff} \twoheadrightarrow \mathbb{P}^{d,n} \tag{4.1}$$

which is an isomorphism over $\mathring{\mathbb{P}}^{d,n}$ (often identified with its inverse image, an open part of $\hat{\mathbb{P}}^{d,n}$).

The canonical stratification of $\hat{\mathbb{A}}^n$ is Aff-stable, hence, induces a stratification $\hat{\mathbb{P}}^{d,n} = \coprod \mathring{\mathbb{P}}^T$. Each stratum $\mathring{\mathbb{P}}^T$ is a connected locally closed subvariety with smooth closure $\mathbb{P}^T$. The strata are parametrized by all connected $n$-groves with at least one vertex. Such a grove will be called an *n-tree* (in [BG] by a tree we meant a connected grove with exactly 2 outgoing edges at each vertex. The latter will be now referred to as a *binary* tree, to conform with the standard terminology). Given a tree $T$ and a vertex $v \in T$, let $|v|$ denote the set of outgoing edges at the vertex $v$. We use the notation $\hat{\mathbb{P}}^{d,|v|}$ instead of $\hat{\mathbb{P}}^{d,\#|v|}$, etc., as in section 2. Write $T' \leq T$ *if the tree $T$ is obtained from $T'$ by contraction of a number of internal edges.*

**Proposition 4.2** (cf. [BG, Prop. 3.7]). (i) $\mathbb{P}^{T'} \subset \mathbb{P}^T$ *if and only if* $T' \leq T$;

(ii) *For any n-tree $T$, there are canonical direct product decompositions:*

$$\mathring{\mathbb{P}}^T \simeq \prod_{\substack{vertices \\ v \in T}} \mathring{\mathbb{P}}^{d,|v|} \qquad \mathbb{P}^T \simeq \prod_{\substack{vertices \\ v \in T}} \hat{\mathbb{P}}^{d,|v|}$$

**Corollary 4.3.** *The closed strata in $\hat{\mathbb{P}}^{d,n}$ are parametrized by binary n-trees. Such a tree $T$ has $n-1$ vertices and we have:* $\mathring{\mathbb{P}}^T = \mathbb{P}^T \simeq (\mathbb{P}^d)^{n-1}$.

Let $T$ be a binary $n$-tree. For any $i, j \geq 0$, the iterated residue map yields a morphism (cf. (3.3)):

$$H^i(\hat{\mathbb{P}}^{d,n}, \Omega^{j+n-2}_{\hat{\mathbb{P}}^{d,n}, \mathring{\mathbb{P}}^{d,n}}) \xrightarrow{Res} H^i(\mathbb{P}^T, \Omega^j_{\mathbb{P}^T}) \simeq H^{i,j}((\mathbb{P}^d)^{n-1}, \mathbb{C}) \tag{4.4}$$

and a similar morphism: $H^{i+n-2}(\mathring{\mathbb{P}}^{d,n}, \mathbb{C}) \to H^i((\mathbb{P}^d)^{n-1}, \mathbb{C})$. The morphisms above depend, up to sign, on the choice of a family of strata $S_1, \ldots, S_k$ as in (3.3). Furthermore, the morphisms corresponding to different binary trees are not all independent. The situation is best described in terms of the Lie algebra language introduced in [BG, §4] and recalled below.

Given a finite set $I$, let $\mathfrak{a}(I)$ denote the Free Lie algebra over $\mathbb{C}$ having the set $I$ as the set of generators. Let $\mathfrak{a}_I \subset \mathfrak{a}(I)$ be the finite dimensional subspace spanned by all bracket expressions that contain each generator once. For the set $I = \{1, \ldots, n\}$ we write $\mathfrak{a}_n$ instead of $\mathfrak{a}_I$. Recall [BG, Lemma 4.2] that there is a natural bijective correspondence between binary trees and certain equivalence classes of formal bracket expressions. We have the following higher dimensional analogue of [BG, Prop. 4.3].

**Proposition 4.5.** *The morphisms (4.4) assembled together (for all binary n-trees) give rise to a canonical vector space isomorphism:*

$$H^i(\hat{\mathbb{P}}^{d,n}, \Omega^{j+n-2}_{\hat{\mathbb{P}}^{d,n}, \mathring{\mathbb{P}}^{d,n}}) \simeq \mathfrak{a}_n^* \otimes H^{i,j}((\mathbb{P}^d)^{n-1}, \mathbb{C}) \qquad, \forall\, i, j \geq 0\,.$$

*Furthermore, the groups on the left vanish if $i < j$ and there are canonical isomorphisms:*

$$H^i(\hat{\mathbb{P}}^{d,n}, \Omega^{i+n-2}_{\hat{\mathbb{P}}^{d,n}, \mathring{\mathbb{P}}^{d,n}}) \simeq H^{2i+n-2}(\hat{\mathbb{P}}^{d,n}, \mathbb{C}) \simeq \mathfrak{a}_n^* \otimes H^{2i}((\mathbb{P}^d)^{n-1}, \mathbb{C}) \qquad, i \geq 0\,.$$

## 5. The Lie Operad and the Resolution of Diagonals

The constructions of this section are special cases of much more general constructions valid for an arbitrary operad (cf. [GK]). A connection between Resolution of Diagonals and the concept of an *operad* was observed implicitly already in [BG]; indeed, the graph complex arising from a spectral sequence for the homology of the Resolution of Diagonals [BG, §6] is nothing but the cobar complex for the Lie operad [GK]. A slightly more general graph complex was introduced, at the same time, by M. Kontsevich in the context of formal non-commutative geometry [Ko]. He then applies that complex for constructing new cohomology classes of various moduli spaces. The non-commutative geometry developed by Kontsevich is part of the Homotopy algebra (going back to earlier works by P. May [Ma], J. Stasheff and others, cf. [HS]), the framework the notion of an operad originally came from. Furthermore, many of Kontsevich's constructions make sense for any Koszul operad (see [GK]). Thus, there is a deep relationship between Operads, Moduli spaces and the Resolution of Diagonals which is not yet fully understood.

To any integer $n \geq 1$ and an $n$-tree $T$, associate a vector space $\mathfrak{a}(T)$ given by the formula, see [BG, (6.3)]:

$$\mathfrak{a}(T) = \bigotimes_{\substack{vertices \\ v \in T}} \mathfrak{a}_{|v|}$$

Let $T$ and $T'$ be a pair of $n$-trees such that $T'$ is obtained from $T$ by contraction of an internel edge $e$. Let $v_s$ be the sourse-vertex and $v_t$ the target vertex of $e$. Observe that $e$ is an element of the set $|v_s|$. Let $v$ be the vertex of $T'$ arising from the contracted edge. Then there is a natural composition morphism (cf. [BG, §§4,6]) $\circ : \mathfrak{a}_{|v_s|} \otimes \mathfrak{a}_{|v_t|} \to \mathfrak{a}_{|v|}$ obtained by inserting the bracket expression formed by elements of $|v_t|$ into the bracket expression formed by elements of $|v_s|$ in place of the generator $e \in |v_s|$. Observe now that the trees $T$ and $T'$ have the same vertices apart from $v_s, v_t$ and $v$. Define a linear map $\mathfrak{a}(T) \to \mathfrak{a}(T')$ to be the tensor product of $\circ$ with the identity morphism on $\bigotimes_{\substack{vertices \\ x \in T', x \neq v}} \mathfrak{a}_{|x|}$. Let, further, a tree $T'$ be obtained from $T$ by contraction of a number of internal edges (cf. Proposition 4.2(i)). Iterating the above construction one defines, via composition, the following natural morphisms:

$$d_{T',T} : \mathfrak{a}(T) \to \mathfrak{a}(T') \quad \text{whenever} \quad T \leq T' \tag{5.1}$$

Observe that the set of all $n$-trees has the unique maximal element with respect to the partial order "$\leq$", the tree $T^\star$ with a single vertex and no internel edges (such a tree was called *star* in [BG]). Clearly, $\mathfrak{a}(T^\star) = \mathfrak{a}_n$. Thus, for any $n$-tree $T$, there is a canonical $\Sigma(T)$-equivariant morphism, arising from (5.1) for $T' = T^\star$:

$$\mathfrak{a}(T) \to \mathfrak{a}_n \tag{5.2}$$

Given an integer $n \geq 1$ and a complex Lie algebra $\mathfrak{g}$, define an *evaluation* map: $\mathfrak{a}_n \otimes \mathfrak{g}^{\otimes n} \to \mathfrak{g}$ by assigning to $a \otimes (\otimes x_i) \in \mathfrak{a}_n \otimes \mathfrak{g}^{\otimes n}$ the element of $\mathfrak{g}$ obtained by inserting the $x_i$'s into the bracket expression given by $a$ in place of the corresponding generators $i = 1, \ldots, n$. The map is clearly invariant under the simultaneous permutation action on $\mathfrak{a}_n$ and on $\mathfrak{g}^{\otimes n}$, hence factors through $(\mathfrak{a}_n \otimes \mathfrak{g}^{\otimes n})_{\Sigma_n}$ where the subscript (resp. superscript) $\Sigma_n$ stands for coinvariants (resp. invariants) of the $\Sigma_n$-action. Let

$$\phi_n : \mathfrak{g}^* \longrightarrow (\mathfrak{g}^{*\otimes n} \otimes \mathfrak{a}_n^*)^{\Sigma_n} \tag{5.3}$$

be the adjoint of the evaluation map. Composing $\phi_n$ with the map $\mathfrak{a}_n^* \to \mathfrak{a}(T)^*$ dual to (5.2), one obtains, for each $n$-tree $T$, a canonical element

$$\Phi(T) \in \mathrm{Hom}\left(\mathfrak{g}^*, \mathfrak{g}^{*\otimes n} \otimes \mathfrak{a}(T)^*\right) = \mathrm{Hom}\left(\mathfrak{a}(T) \otimes \mathfrak{g}^{\otimes n}, \mathfrak{g}\right) \tag{5.4}$$

The element $\Phi(T)$ is invariant under the natural $\Sigma(T)$-action.

Given a finite set $I$, let $U(I)$ denote the free associative $\mathbb{C}$-algebra having $I$ as the set of generators, and let $\mathfrak{A}_I$, denote the subspace of $U(I)$ spanned by the monomials that contain each generator once (put also $\mathfrak{A}_\emptyset = \mathbb{C}$). The algebra $U(I)$ may be identified with the enveloping algebra of the free Lie algebra $\mathfrak{a}(I)$ so that there is a natural imbedding $\mathfrak{a}_I \hookrightarrow \mathfrak{A}_I$.

Let $e \in I$ be a marked element of a finite set $I$. An $I$-tree will be referred to as a *marked* tree, to be denoted $\overset{\bullet}{T}$. The oriented path starting from the externel ingoing edge of the tree and ending with the externel outgoing edge marked by the element $e$ is called the marked string of the tree $\overset{\bullet}{T}$. The vertices and the edges that belong to the marked string are said to be *marked*, the others are said to be *non-marked*. Let $\Sigma(\overset{\bullet}{T}) \subset \Sigma_I$ denote the subgroup of the permutation group of the set $I$ that fixes the isomorphism class of the tree as well as its marked externel edge $e$.

We can now repeat all the above constructions for marked trees.

First, to any marked $n$-tree $\overset{\bullet}{T}$, associate the vector space:

$$\mathfrak{A}(\overset{\bullet}{T}) = \Big(\bigotimes_{\substack{marked \\ vertices}} \mathfrak{A}_{|v|}\Big) \otimes \Big(\bigotimes_{\substack{nonmarked \\ vertices}} \mathfrak{a}_{|v|}\Big)$$

Next, let $\overset{\bullet}{T}$ and $\overset{\bullet}{T}'$ be a pair of marked trees such that $\overset{\bullet}{T}'$ is obtained from $\overset{\bullet}{T}$ by contraction of an internel edge $e$ with sourse-vertex $v_s$ and the target vertex $v_t$. We wish to construct a morphism $\mathfrak{A}(\overset{\bullet}{T}) \to \mathfrak{A}(\overset{\bullet}{T}')$. There are several different cases. Assume first that both endpoints of the edge $e$ are nonmarked. Then the vertex $v \in \overset{\bullet}{T}'$ arising from the contracted edge is nonmarked also, so that there is the composition morphism $\circ : \mathfrak{a}_{|v_s|} \otimes \mathfrak{a}_{|v_t|} \to \mathfrak{a}_{|v|}$ used already in the construction of morphisms (5.1). The trees $\overset{\bullet}{T}$ and $\overset{\bullet}{T}'$ have the same vertices, apart from $v_s$, $v_t$ and $v$. So, one defines the map $\mathfrak{A}(\overset{\bullet}{T}) \to \mathfrak{A}(\overset{\bullet}{T}')$ to be the tensor product of " $\circ$ " with the identity maps on other factors. Assume next that the vertex $v_t$ is marked. Then the edge $e$, hence the vertex $v_s$, are also marked. In such a case one defines a similar composition map $\circ : \mathfrak{A}_{|v_s|} \otimes \mathfrak{A}_{|v_t|} \to \mathfrak{A}_{|v|}$, and procedes as in the previous case. Assume finally that the vertex $v_t$ is nonmarked while the vertex $v_s$ is marked. Then the resulting vertex $v$ is marked. Combine the natural imbedding $\mathfrak{a}_{|v_t|} \hookrightarrow \mathfrak{A}_{|v_t|}$ with the composition morphism $\mathfrak{A}_{|v_s|} \otimes \mathfrak{A}_{|v_t|} \to \mathfrak{A}_{|v|}$ to obtain a morphism $\mathfrak{A}_{|v_s|} \otimes \mathfrak{a}_{|v_t|} \to \mathfrak{A}_{|v|}$. Again, define a map $\mathfrak{A}(\overset{\bullet}{T}) \to \mathfrak{A}(\overset{\bullet}{T}')$ to be the product of that morphism with the identity morphisms on the other factors. Thus, we have constructed the map $\mathfrak{A}(\overset{\bullet}{T}) \to \mathfrak{A}(\overset{\bullet}{T}')$ in all cases. Iterating the above construction one defines (cf. (5.1)), via composition, natural morphisms :

$$\mathfrak{A}(\overset{\bullet}{T}) \to \mathfrak{A}(\overset{\bullet}{T}') \quad \text{whenever} \quad \overset{\bullet}{T} \leq \overset{\bullet}{T}' \tag{5.5}$$

Now let $\mathfrak{g}$ be a Lie algebra and $U(\mathfrak{g})$ its enveloping algebra. For any $n \geq 1$, there is a natural map $\mathfrak{A}_n \otimes \mathfrak{g}^{\otimes n} \to U(\mathfrak{g})$ assigning to $u \otimes (\otimes x_i)$, $u \in \mathfrak{A}_n, x_i \in \mathfrak{g}$, the element $u(x) \in U(\mathfrak{g})$ obtained by inserting the $x_i$'s into $u$ instead of the corresponding generators $1, \ldots, n$. Given a finite dimensional $\mathfrak{g}$-module $E$, define an evaluation map $\mathfrak{A}_n \otimes \mathfrak{g}^{\otimes n} \otimes E \to E$ by sending $u \otimes (\otimes x_i) \otimes v$ to $u(x) \cdot v \in E$. Let

$$\psi_n : E^* \to E^* \otimes \mathfrak{g}^{*\otimes n} \otimes \mathfrak{A}_n^* \tag{5.6}$$

be the adjoint to the evaluation map. Composing it with morphism (5.5) for $\overset{\bullet}{T}' = T^\star$ (=star) one obtains, for any marked $n$-tree $\overset{\bullet}{T}$ and any $\mathfrak{g}$-module $E$, a canonical morphism

$$\Psi(\overset{\bullet}{T}) : E^* \to E^* \otimes \left( \mathfrak{g}^{*\otimes(n-1)} \otimes \mathfrak{A}(\overset{\bullet}{T})^* \right)^{\Sigma(\overset{\bullet}{T})} \tag{5.7}$$

Given a marked tree $\overset{\bullet}{T}$ we may view it as a non-marked tree $T$. The spaces $\mathfrak{a}(T)$ and $\mathfrak{A}(\overset{\bullet}{T})$ are then become canonically isomorphic. To see this, for any finite set $I$ with marked element $e$, put $\overset{\bullet}{I} = I \setminus \{e\}$. Define a linear map $\epsilon : \mathfrak{A}_{\overset{\bullet}{I}} \to \mathfrak{a}_I$ by assigning to a monomial $i_1 \cdot i_2 \cdot \ldots \cdot i_r \in \mathfrak{A}_{\overset{\bullet}{I}}$ the bracket expression $[i_1, [i_2, \ldots, [i_r, e] \ldots] \in \mathfrak{a}_I$. (The assignment arises from the adjoint action of the enveloping algebra $U(\mathfrak{a}(I))$ on $\mathfrak{a}(I)$ ). The map $\epsilon$ turns out to be a bijection; in particular, $\dim \mathfrak{a}_I = \dim \mathfrak{A}_{\overset{\bullet}{I}} = (\#I - 1)!$, as is classically known. It follows that there is a canonical isomorphism $\mathfrak{a}(T) \simeq \mathfrak{A}(\overset{\bullet}{T})$.

We now relate the algebraic constructions above to the geometry of the resolution of diagonals $\hat{\mathbb{P}}^{d,n}$. A sheaf $\mathcal{F}$ on $\hat{\mathbb{P}}^{d,n}$ is said to be *constructible* if its restriction to any stratum $\overset{\circ}{\mathbb{P}}^T$ is a constant sheaf with fiber $\mathcal{F}_T$. The assingment $\mathcal{F} \rightsquigarrow \{\mathcal{F}_T\}$ sets up an equivalence of the category of constructible sheaves on $\hat{\mathbb{P}}^{d,n}$ with the category whose objects are collections of finite dimensional vector spaces $\{\mathcal{F}_T\}$, one for each $n$-tree $T$, equipped, for every pair $T' \leq T$ (cf. Proposition 4.2(i)) with transition morphisms between the dual spaces $Res_{T,T'} : \mathcal{F}_T^* \to \mathcal{F}_{T'}^*$. The transition maps should satisfy natural compatibility conditions for every triple $T'' \leq T' \leq T$. For example, the collection $\mathfrak{a} = \{\mathfrak{a}(T)\}$ (more generally, any operad) gives a $\Sigma_n$-equivariant constructible sheaf $\underline{\mathfrak{a}}$ on $\hat{\mathbb{P}}^{d,n}$. The transition morphisms in this example are given by maps (5.1).

Given a Lie algebra $\mathfrak{g}$, let $\mathfrak{g}_{\hat{\mathbb{P}}}$ denote the constant sheaf on $\hat{\mathbb{P}}^{d,n}$ with fiber $\mathfrak{g}$.

**Proposition 5.8.** *The collection $\{\Phi(T)\}$, see (5.2), gives rise to a $\Sigma_n$-equivariant sheaf morphism: $\mathfrak{g}_{\hat{\mathbb{P}}}^{\otimes n} \otimes \underline{\mathfrak{a}} \to \mathfrak{g}_{\hat{\mathbb{P}}}$, in other words one has:*
$$Res_{T,T'} \Phi(T) = \Phi(T').$$

Similar result holds for the collection $\{\Psi(\overset{\bullet}{T})\}$.

There is a more deep relationship between resolution of diagonals and Free Lie algebras, arising from Proposition 4.5. The isomorphism of the Proposition shows that its left-hand side vanishes whenever $i > d(n-1)$ or $j > d \cdot n - d - 1$. Moreover, we have:
$$H^{2d(n-1)}(\overset{\circ}{\mathbb{P}}^{d,n}, \mathbb{C}) \simeq H^{d(n-1)}(\hat{\mathbb{P}}^{d,n}, \Omega_{\mathbb{P}^{d,n}, \overset{\circ}{\mathbb{P}}^{d,n}}) \simeq \mathfrak{a}_n^*$$

For any $n$-tree $T$, set $\Omega_T = \Omega_{\mathbb{P}^T, \mathring{\mathbb{P}}^T}$. Observe that there is a trivial identity: $n - 1 = \sum_{\substack{vertices \\ v \in T}} (\#|v| - 1)$. Whence, by Proposition 4.2 and the Kunneth formula, one obtains the following result.

**Lemma 5.9.** *The groups $H^i(\mathbb{P}^T, \Omega_T)$ vanish for all $i > d(n-1)$, and there are canonical isomorphisms:*

$$H^{2d(n-1)}(\mathbb{P}^T, \mathbb{C}) \simeq H^{d(n-1)}(\mathbb{P}^T, \Omega_T) \simeq \mathfrak{a}(T)^* \ (resp. \simeq \mathfrak{A}(\mathring{T})^*)$$

The above described transition maps for the sheaf $\underline{\mathfrak{a}}$ on $\hat{\mathbb{P}}^{d,n}$ are nothing but the iterated residue maps: $H^{d(n-1)}(\mathbb{P}^T, \Omega_T) \xrightarrow{res} H^{d(n-1)}(\mathbb{P}^{T'}, \Omega_{T'})$. Furthermore, the maps (5.2) and (5.7) may be viewed as morphisms

$$\begin{aligned} \Phi(T) : \mathfrak{g}^* &\longrightarrow \mathfrak{g}^{*\otimes n} \otimes H^{d(n-1)}(\mathbb{P}^T, \Omega_T) \\ \Psi(\mathring{T}) : E^* &\longrightarrow E^* \otimes \mathfrak{g}^{*\otimes(n-1)} \otimes H^{d(n-1)}(\mathbb{P}^{\mathring{T}}, \Omega_{\mathring{T}}) \end{aligned} \tag{5.10}$$

## 6. Sheaves on the Resolution of Diagonals

Let $X$ be a smooth $d+1$-dimensional variety. For any integers $n \geq 1$ and $m \geq 0$, there is a natural projection $p_m : \hat{X}^{m+n} \twoheadrightarrow X^{m+n} \twoheadrightarrow X^m$. Let $x_1, \ldots, x_m$ be a fixed (possibly empty) collection of pairwise distinct points of $X$, and $\hat{X}_m^n := p_m^{-1}(x_1, \ldots, x_m)$ the fiber of $p_m$ over the point $(x_1, \ldots, x_m) \in \mathring{X}^m$ (so that $\mathring{X}^m = pt$ and $\hat{X}_m^n = \hat{X}^n$ if $m = 0$). The map $p_m$ being strata preserving, it follows that $\hat{X}_m^n$ is a smooth $n$-dimensional variety with Zariski open part $\mathring{X}_m^n$, the fiber over $(x_1, \ldots, x_m)$ of the projection $\mathring{X}^{m+n} \twoheadrightarrow \mathring{X}^m$. Furthermore, $\hat{X}_m^n \setminus \mathring{X}_m^n$ is a normal crossing divisor with smooth irreducible components. The components induce a stratification $\hat{X}_m^n = \coprod_T S_T$ (which coincides with the stratification induced from that on $\hat{X}^{m+n}$) with the properties described in Proposition 6.2 below.

Let $T$ be an $(m+n)$-grove with $m+r$ connected components $T_1, \ldots, T_m, \ldots, T_{m+r}$, $r \geq 0$. Put:

$$\mathring{\mathbb{P}}^T := \prod_{i=1}^{m+r} \mathring{\mathbb{P}}^{T_i} \qquad (\text{ resp. } \mathbb{P}^T := \prod_{i=1}^{m+r} \mathbb{P}^{T_i})$$

**Definition 6.1.** The grove $T$ is said to be an $(m,n)$-*grove* if, for any $i = 1,\ldots,m$, the integer $i$ labeles an externel edge of the component $T_i$. The components $T_1,\ldots,T_m$ are viewed as marked trees with marked edges labeled by $1,\ldots,m$, respectively (so that any connected component has at most one marked edge).

**Proposition 6.2** (cf. [BG, Prop. 3.6]). *(i) The strata $S_T$ of the stratification of $\hat{X}_m^n$ are parametrized by all $(m,n)$-groves $T$;*

*(ii) The stratum $S_T$ (resp. its closure) is isomorphic canonically to an algebraic fibration:*

$$S_T \xrightarrow{\mathring{\mathbb{P}}^T} \mathring{X}_m^r \qquad (resp.\ \bar{S}_T \xrightarrow{\mathbb{P}^T} \hat{X}_m^r)$$

*with fiber $\mathring{\mathbb{P}}^T$ (resp. $\mathbb{P}^T$), where $m+r$ is the number of connected components of the grove $T$.*

Given an $(m,n)$-grove $T$, let $\pi_T : \bar{S}_T \to \hat{X}_m^r$ be the canonical fibration and $\Omega_{rel}$ the sheaf of relative top degree forms on the fibration with logarithmic singularities at the boundary. It follows from Lemma 5.6 and Proposition 6.1(ii) that the higher direct image sheaves $(\mathcal{H}^i\pi_T)_*\Omega_{rel}$ vanish for $i > d(n-r)$ and $(\mathcal{H}^{d(n-r)}\pi_T)_*\Omega_{rel}$ is a free $\mathcal{O}$-sheaf on $\hat{X}_m^r$ whose fiber is canonically isomorphic to:

$$(\mathcal{H}^{d(n-r)}\pi_T)_*\Omega_{rel} \simeq H^{d(n-r)}(\mathbb{P}^T, \Omega_T) \tag{6.3}$$

Observe further that there is a natural commutative diagram:

$$\begin{array}{ccc} \bar{S}_T & \hookrightarrow & \hat{X}_m^n \\ \pi_T \downarrow & & \downarrow \pi \\ \hat{X}_m^r \to X^r & \xrightarrow{i} & X^n \end{array} \tag{6.4}$$

where $i$ stands for the imbedding of $X^r$ as the diagonal of $X^n$ formed by points whose coordinates are set equal whenever the corresponding labels of the grove $T$ belong to the same connected component.

Let $G$ be a complex Lie group with Lie algebra $\mathfrak{g}$ and $P$ a principal holomorpic $G$-bundle on the variety $X$. Given a finite dimensional $G$-module $E$, let $E_P$ denote the associated vector bundle on $X$. In particular, $\mathfrak{g}_P$, resp. $\mathfrak{g}_P^*$, denotes the associated vector bundle corresponding to the adjoint, resp. coadjoint, representation. From now on, fix a finite collection of finite dimensional $G$-modules $E_1, \ldots, E_m$, one for each point $x_i$, and let $\mathcal{E}_1, \ldots, \mathcal{E}_m$ denote the associated vector bundles on $X$.

Form the externel tensor product $\mathcal{G}_m^n := \mathcal{E}_1^* \boxtimes \ldots \boxtimes \mathcal{E}_m^* \boxtimes \mathfrak{g}_P^{*\boxtimes n}$, a locally free sheaf on $X^{m+n}$. Let $\hat{\mathcal{G}}_m^n$ denote its pull-back to $\hat{X}_m^n$ via the composition of the natural maps:

$$\hat{X}_m^n \hookrightarrow \hat{X}^{m+n} \twoheadrightarrow X^{m+n}$$

Let $S_T$ be the stratum associated to an $(m,n)$-grove $T$ with $m+r$ components. The sheaf $\hat{\mathcal{G}}_{m\,|\bar{S}_T}^n$, the restriction of $\hat{\mathcal{G}}_m^n$ to $\bar{S}_T$, is constant along the fibers of the projection $\pi_T$ (see (6.4)), moreover, we have: $\hat{\mathcal{G}}_{m\,|\bar{S}_T}^n = \pi_T^*(j^*\mathcal{G}_m^n)$, where $j$ stands for the composition of the maps in the low row of diagram (6.4). Set, cf. (5.7) and the Kunneth formula:

$$\Psi(T) = \bigotimes_{i=1}^m \Psi(\overset{\bullet}{T}_i) \otimes \bigotimes_{i=m+1}^{m+r} \Phi(T_i) \;:$$

$$(\otimes_i E_i^*) \otimes \mathfrak{g}^{*\otimes r} \to (\otimes_i E_i^*) \otimes \mathfrak{g}^{*\otimes n} \otimes H^{d(n-r)}(\mathbb{P}^T, \Omega_T)$$

Thus, formula (6.3) combined with the projection formula shows that the map $\Psi(T)$ gives rise to a canonical morphism of sheaves on $\hat{X}_m^r$:

$$\hat{\mathcal{G}}_m^r \otimes \Omega_{\hat{X}_m^r, \overset{\circ}{X}_m^r} \to (\mathcal{H}^{d(n-r)} \pi_T)_* (\hat{\mathcal{G}}_{m\,|\bar{S}_T}^n \otimes \Omega_{\bar{S}_T, S_T}) \qquad (6.5)$$

This morphism is a locally split imbedding, provided $\mathfrak{g}$ is a semisimple Lie algebra.

## 7. Jets on Moduli spaces of G-bundles

Let $G$ be a complex semisimple Lie group with Lie algebra $\mathfrak{g}$ and $\mathcal{M}$ the moduli space of *stable* holomorphic (equivalently, algebraic) principal $G$-bundles on a smooth compact algebraic variety $X$ of dimension $d+1$ (the condition of stability may be dropped if $\dim X = 1$). Given a point $x \in X$, let $\mathcal{P}_{\mathcal{M},x}$ denote the principal $G$-bundle on $\mathcal{M}$ whose fiber at each point $P \in \mathcal{M}$ is the fiber of the bundle $P$ at $x$. We keep the notation of §6. In particular, we fix a finite collection of points $x_1, \ldots, x_m \in X$ and the same number of finite dimensional $G$-modules $E_1, \ldots, E_m$. Let $\mathcal{E}_{\mathcal{M},x_i}$ denote the associated vector bunle on $\mathcal{M}$ corresponding to the $G$-module $E_i$ and the principal $G$-bundle $\mathcal{P}_{\mathcal{M},x_i}$. Put $\mathcal{E}_\mathcal{M} = \otimes_i \mathcal{E}_{\mathcal{M},x_i}$ and let $\mathfrak{J}_P^n(\mathcal{E}_\mathcal{M}^*)$ denote the vector space of $n$-jets at $P \in \mathcal{M}$ of the sheaf of regular sections of the dual bundle $\mathcal{E}_\mathcal{M}^*$. The aim of this section is to give, for each $n \geq 0$, a canonical construction of the vector spaces $\mathfrak{J}_P^n(\mathcal{E}_\mathcal{M}^*)$, a generalization of the main result of [BG].

If $n = 0$ we have $\hat{X}_m^0 = pt$, and there is an obvious isomorphism $\mathfrak{J}_P^0(\mathcal{E}_\mathcal{M}^*) = \mathcal{G}_m^0$. For $n = 1$ and any $P \in \mathcal{M}$, the Kodaira-Spencer formula combined with Serre duality yields a short exact sequence:

$$0 \to \mathcal{G}_m^0 \otimes H^d(X, \mathfrak{g}_P^* \otimes \Omega_X) \to \mathfrak{J}_P^1(\mathcal{E}_\mathcal{M}^*) \to \mathfrak{J}_P^0(\mathcal{E}_\mathcal{M}^*) \to 0 \qquad (7.1)$$

The Theorem below shows, in the special case $n = 1$, that (7.1) is isomorphic canonically to a natural short exact sequence:

$$0 \to \mathcal{G}_m^0 \otimes H^d(X, \mathfrak{g}_P^* \otimes \Omega_X) \to H^d(X, \mathcal{G}_m^1 \otimes \Omega_X) \to \mathcal{G}_m^0 \to 0$$

Now let $n \geq 1$. If $P \in \mathcal{M}$ is a stable $G$-bundle on $X$ then the vector bundle $\mathfrak{g}_P$ has no global sections. Hence, Serre duality yields:

$$H^{d+1}(X, \mathfrak{g}_P^* \otimes \Omega_X) = 0 \qquad , \quad P \in \mathcal{M}$$

From this cohomology vanishing and the Kunneth formula, one proves, by induction on $n$, the following result.

**Lemma 7.2.** *For any stable $G$-bundle $P$ we have:*

$$H^i(\hat{X}_m^n, \hat{\mathcal{G}}_m^n \otimes \Omega_{\hat{X}_m^n, \overset{\circ}{X}_m^n}) = 0 \quad \text{if} \quad i > d \cdot n \quad .$$

Given an $(m,n)$-grove $T$, let $Res_T : H^*(\hat{X}_m^n, \hat{\mathcal{G}}_m^n \otimes \Omega_{\hat{X}_m^n, \mathring{X}_m^n}) \to H^*(\bar{S}_T, \Omega_{\bar{S}_T, S_T})$ be the corresponding iterated residue map.

Definition. $H^{\spadesuit}(\hat{X}_m^n, \hat{\mathcal{G}}_m^n \otimes \Omega_{\hat{X}_m^n, \mathring{X}_m^n}) := \{\omega \in H^{d \cdot n}(\hat{X}_m^n, \hat{\mathcal{G}}_m^n \otimes \Omega_{\hat{X}_m^n, \mathring{X}_m^n})$ such that, $\omega$ is anti-invariant with respect to the action of the subgroup $Id_m \times \Sigma_n \subset \Sigma_{m+n}$ and, for every $(m,n)$-grove $T$, the element $Res_T \omega$ satisfies the following condition:

$Res_T \omega$ is the image of a $\dot{\Sigma}_{r+m}$-anti-symmetric section under the natural morpism

$$H^{d \cdot r}(\hat{X}_m^r, \hat{\mathcal{G}}_m^r \otimes \Omega_{\hat{X}_m^r, \mathring{X}_m^r}) \xrightarrow{(6.5)}$$

$$\to H^{d \cdot r}\left(\hat{X}_m^r, (\mathcal{H}^{d(n-r)} \pi_T)_* (\hat{\mathcal{G}}_{m|\bar{S}_T}^n \otimes \Omega_{\bar{S}_T, S_T})\right) \simeq H^{d \cdot n}(\bar{S}_T, \hat{\mathcal{G}}_m^n \otimes \Omega_{\bar{S}_T, S_T})$$

Here $\dot{\Sigma}_{r+m}$ stands for the subgroup of the isotropy group $\Sigma(T)$ that keeps all the labels $1, \ldots, m$ fixed, and the last isomorphism is due to Lemma 7.2. Observe that if $\dim X = d+1 = 1$ so that $d = 0$, then the latter isomorphism holds without any stability assumptions on $P$.

Here is the main result of this section (cf. [BG, Thm. 8.3]).

Theorem 7.3. *For any $P \in \mathcal{M}$ and $n \geq 1$ there is a canonical isomorphism of vector spaces:*

$$\mathfrak{J}_P^n(\mathcal{E}_{\mathcal{M}}^*) \simeq H^{\spadesuit}(\hat{X}_m^n, \hat{\mathcal{G}}_m^n \otimes \Omega_{\hat{X}_m^n, \mathring{X}_m^n}) \quad .$$

Remark 7.4. Given two collections of points $\{'x_1, \ldots, 'x_m\}$ and $\{''x_1, \ldots, ''x_l\}$, and two collections of finite dimensional $G$-modules $\{'E_1, \ldots, 'E_m\}$ and $\{''E_1, \ldots, ''E_l\}$, let $'\mathcal{E}_{\mathcal{M}}^*$ and $''\mathcal{E}_{\mathcal{M}}^*$ denote the corresponding vector bundles on the moduli space. Then there is a natural multiplication of jet spaces:

$$\mathfrak{J}_P^n('\mathcal{E}_{\mathcal{M}}^*) \otimes \mathfrak{J}_P^k(''\mathcal{E}_{\mathcal{M}}^*) \to \mathfrak{J}_P^{n+k}('\mathcal{E}_{\mathcal{M}}^* \otimes ''\mathcal{E}_{\mathcal{M}}^*)$$

This multiplication corresponds, via the isomorphism of the Theorem and the Kunneth formula, to the composition map (where $q = d(n+k)$) :

$$\mathfrak{J}_P^n({}'\mathcal{E}_\mathcal{M}^*) \otimes \mathfrak{J}_P^k({}''\mathcal{E}_\mathcal{M}^*) \simeq H^q\Big(\hat{X}_m^n \times \hat{X}_l^k,\, (\hat{\mathcal{G}}_m^n \boxtimes \hat{\mathcal{G}}_l^k) \otimes (\Omega_{\hat{X}_m^n, \mathring{X}_m^n} \boxtimes \Omega_{\hat{X}_l^k, \mathring{X}_l^k})\Big) \xrightarrow{\sim}$$

$$H^q\Big(\hat{X}_{m+l}^{n+k},\, \widehat{(\mathcal{G}_m^n \otimes \mathcal{G}_l^k)} \otimes \Omega_{\hat{X}_{m+l}^{n+k}, \mathring{X}_m^n \times \mathring{X}_l^k}\Big) \to H^q\Big(\hat{X}_{m+l}^{n+k},\, \hat{\mathcal{G}}_{m+l}^{n+k} \otimes \Omega_{\hat{X}_{m+l}^{n+k}, \mathring{X}_{m+l}^{n+k}}\Big)$$

Representation theoretic interpretation: Let $d = 1$ so that $X$ is a complex compact curve. For each point $x_i \in X$, $i = 1, \ldots, m$, let $\mathcal{O}_i$ denote the local ring at $x_i$, $\mathfrak{M}_i \subset \mathcal{O}_i$ its maximal ideal, and $\mathfrak{R}_i$ the field of fractions of the ring $\mathcal{O}_i$. The tensor product $\mathfrak{g}(\mathfrak{R}_i) := \mathfrak{g} \otimes_\mathbb{C} \mathfrak{R}_i$ has a natural Lie algebra structure, the *Loop algebra* at the point $x_i$. Using similar notation, we have a canonical short exact sequence of Lie algebras $\mathfrak{g}(\mathfrak{M}_i) \hookrightarrow \mathfrak{g}(\mathcal{O}_i) \twoheadrightarrow \mathfrak{g}$. Given a finite dimensional $\mathfrak{g}$-module $E_i$, pull it back to $\mathfrak{g}(\mathcal{O}_i)$ and let $M(E_i) = U(\mathfrak{g}(\mathfrak{R}_i)) \otimes_{U(\mathfrak{g}(\mathcal{O}_i))} E_i$ denote the "partial" Verma module induced from $E_i$. It has a natural increasing filtration: $M_0(E_i) \subset M_1(E_i) \subset \ldots$ induced by the standard increasing filtration on the enveloping algebra. Form the direct product of Lie algebras $\mathbf{L}\mathfrak{g} = \mathfrak{g}(\mathfrak{R}_1) \times \ldots \times \mathfrak{g}(\mathfrak{R}_m)$, and let $\mathbb{M} = M(E_1) \otimes \ldots \otimes M(E_m)$ be the corresponding Verma module over $\mathbf{L}\mathfrak{g}$, equipped with the tensor product filtration $\mathbb{M}_1 \subset \mathbb{M}_2 \subset \ldots$.

Now, let $X_{out} := X \setminus \{x_1, \ldots, x_m\}$ be the complement of the points, an affine curve with regular ring $\mathcal{O}(X_{out})$. There is a natural restriction homomorphism: $\mathcal{O}(X_{out}) \hookrightarrow \oplus_{1 \leq i \leq m} \mathfrak{R}_i$ which is clearly injective. It gives rise to the corresponding injective Lie algebra homomorphism:

$$\mathfrak{g}_{out} := \mathfrak{g} \otimes_\mathbb{C} \mathcal{O}(X_{out}) \hookrightarrow \oplus_{1 \leq i \leq m} \mathfrak{g}(\mathfrak{R}_i) = \mathbf{L}\mathfrak{g}$$

Let $H_0(\mathfrak{g}_{out}, \mathbb{M}) = \mathbb{M}/\mathfrak{g}_{out} \cdot \mathbb{M}$ be the space of *coinvariants* of the $\mathfrak{g}_{out}$-module $\mathbb{M}$. The filtration on $\mathbb{M}$ induces, via the projection $\mathbb{M} \twoheadrightarrow \mathbb{M}/\mathfrak{g}_{out} \cdot \mathbb{M}$, a natural filtration $H_0(\mathfrak{g}_{out}, \mathbb{M})_n$, $n = 1, 2, \ldots$ on the coinvariants.

Theorem 7.3 is equivalent to the following result which is proved by an argument similar (and essentially equivalent) to the argument used [BD].

Theorem 7.3'. *For any $n \geq 1$, there is a natural vector space isomorphism:*

$$\Big(H_0(\mathfrak{g}_{out}, \mathbb{M})_n\Big)^* \simeq H^\spadesuit(\hat{X}_m^n, \hat{\mathcal{G}}_m^n \otimes \Omega_{\hat{X}_m^n, \mathring{X}_m^n}) \qquad \square$$

**Moduli spaces of $G$-bundles with flat connection:** There is a topological counterpart of Theorem 7.3 for a moduli space of $G$-bundles with flat connection. Let $X$ be a compact Kahler manifold and $\hat{\mathcal{M}}$ the set of all homomorphisms $\gamma : \pi_1(X) \to G$, where $\pi_1(X)$ denotes the fundamental group of $X$ with respect to a fixed point $x \in X$. Let $\mathcal{M}'$ be the subset of $\hat{\mathcal{M}}$ formed by all the homomorphisms $\gamma$ that have the following property: *the centralizer in $G$ of the image of $\gamma$ equals the center of $G$.* The $G$-conjugacy classes of homomorphisms $\gamma \in \mathcal{M}'$ form, in a natural way, a smooth stack $\mathcal{M}^\nabla$. Pulling back the adjoint $G$-action on $\mathfrak{g}$ via a homomorphism $\gamma \in \mathcal{M}^\nabla$ makes the space $\mathfrak{g}$ a into $\pi_1(X)$-module, $\mathfrak{g}_\gamma$, and one has:

$$H^0(\pi_1(X), \mathfrak{g}_\gamma) = 0 \quad , \quad H^1(\pi_1(X), \mathfrak{g}_\gamma) = T_\gamma(\mathcal{M}^\nabla)$$

Further, giving a homorphism $\pi_1(X) \to G$ is the same as giving a principal $C^\infty$-bundle on $X$ with flat connection. Thus, the stack $\mathcal{M}^\nabla$ may be (and will be) identified with the moduli space of $C^\infty$-bundle with flat connection. A flat connection on a $G$-bundle $P$ induces flat connections on all the associated vector bundles. Thus, given a collection of points of $X$ and a collection of $G$-modules, one obtains in a natural way the following local systems: $\mathfrak{g}_P$, $\mathfrak{g}_P^*$, $\mathcal{G}_m^n$, $\hat{\mathcal{G}}_m^n$ that have been defined previously in the context of holomorphic vector bundles.

Next, let $T$ be an $(m, n)$-grove, $S_T$ the corresponding stratum in $\hat{X}_{\mathfrak{g}^n}^n$, $\mathbb{C}_T$ the constant sheaf on $S_T$, and $(\mathcal{H}^i \pi_T)_* \mathbb{C}_T$ its higher direct image to $\mathring{X}_m^r$ via the canonical projection (cf. Proposition 6.2) $\pi_T : S_T \to \mathring{X}_m^r$. Lemma 5.6 yields the following analogue of isomorphism (6.3):

$$(\mathcal{H}^{2d(n-1)} \pi_T)_* \mathbb{C}_T \simeq H^{2d(n-1)}(\mathbb{P}^T, \mathbb{C})$$

One can now repeat all the constructions of §6 with coherent sheaves being replaced by local systems to get a canonical morphism : $\hat{\mathcal{G}}_m^r \to$
$(\mathcal{H}^{2d(n-r)} \pi_T)_* (\hat{\mathcal{G}}_{m|S_T}^n)$ , which is a topological analogue of morphism (6.5).

**Definition.** $H^\nabla(\mathring{X}_m^n, \hat{\mathcal{G}}_m^n) := \{\omega \in H^{2d \cdot n}(\mathring{X}_m^n, \hat{\mathcal{G}}_m^n)$ such that $\omega$ is anti-invariant with respect to the action of the subgroup $Id_m \times \Sigma_n \subset \Sigma_{m+n}$ and, for every $(m, n)$-grove $T$, the element $Res_T \omega$ satisfies the following condition:

$Res_T \omega$ is the image of a $\overset{\bullet}{\Sigma}_{r+m}$-anti-symmetric section under the natural morpism

$$H^{2d\cdot r}(\mathring{X}_m^r, \hat{\mathcal{G}}_m^r) \to H^{2d\cdot r}\left(\mathring{X}_m^r, (\mathcal{H}^{2d(n-r)}\pi_T)_* \hat{\mathcal{G}}_{m|S_T}^n\right) \simeq H^{2d\cdot n}(S_T, \hat{\mathcal{G}}_m^n)$$

□

Further, there is a natural vector bundle $\mathcal{E}_\mathcal{M}$ on $\mathcal{M}^\nabla$, similar to the one on $\mathcal{M}$, and we have the following analogue of Theorem 7.3.

**Theorem 7.5.** *For any $P \in \mathcal{M}^\nabla$ and $n \geq 1$, there is a canonical isomorphism of the vector spaces, $\mathfrak{J}_P^n(\mathcal{E}_\mathcal{M}^*)$, of jets on $\mathcal{M}^\nabla$:*

$$\mathfrak{J}_P^n(\mathcal{E}_\mathcal{M}^*) \simeq H^\nabla(\mathring{X}_m^n, \hat{\mathcal{G}}_m^n)$$

## 8. Quasi-Lie algebras and central extensions

We introduce here the notion of a *quasi-Lie algebra*, a generalization of Lie algebras with a slightly violated Jacobi identity. Interesting examples of such objects arise naturally from the central extensions of loop algebras. These objects will play an important role in the description of jet spaces of sections of determinant bundles outlined in the next section.

Given a vector space $\tilde{\mathfrak{g}}$ and a skew-symmetric bilinear map $[\,,\,]: \tilde{\mathfrak{g}} \times \tilde{\mathfrak{g}} \to \tilde{\mathfrak{g}}$, set:

$$J(x, y, z) = [x, [y, z]] + [z, [x, y]] + [y, [z, x]] \tag{8.1}$$

Observe that this expression enters the Jacobi identity, and that (8.1) defines a morphism $J: \Lambda^3 \tilde{\mathfrak{g}} \to \tilde{\mathfrak{g}}$. Now let the space $\tilde{\mathfrak{g}}$ with skew-symmetric bracket $[\,,\,]$ be equipped, in addition, with a vector subspace $C \subset \tilde{\mathfrak{g}}$.

**Definition 8.2.** *The triple $(\tilde{\mathfrak{g}}, C, [\,,\,])$ is called a quasi-Lie algebra if the following holds:*

  *(i)  The subspace $C$ is central with respect to the bracket, i.e.,*
      $[C, \tilde{\mathfrak{g}}] \equiv 0$;

(ii) *The bracket $[\,,\,]$ induces a Lie algebra structure on the quotient $\tilde{\mathfrak{g}}/C$ , that is $J(x,y,z) \in C$ , for any $x,y,z \in \tilde{\mathfrak{g}}$ .*

**Example 8.3.** Let $\mathfrak{g}$ be a complex Lie algebra with a symmetric invariant bilinear form $\langle\,,\,\rangle : S^2\mathfrak{g} \to \mathbb{C}$. Let $A = \mathbb{C}[\varepsilon]/(\varepsilon^2)$ denote the ring of Dual numbers and $\mathfrak{g}[A] := A \otimes_{\mathbb{C}} \mathfrak{g}$ , the Lie algebra over $A$ obtained by extension of scalars. Set $\tilde{\mathfrak{g}} := \mathfrak{g}[A] \oplus \mathbb{C}$ and $C := \mathbb{C} \subset \tilde{\mathfrak{g}}$ , and define a bracket on $\tilde{\mathfrak{g}}$ by the formula:

$$[x + \varepsilon x' \oplus a \,,\, y + \varepsilon y' \oplus b] = [x,y] + \varepsilon([x,y'] + [x',y]) \oplus (\langle x,y'\rangle - \langle x',y\rangle)$$

where $x, x', y, y' \in \mathfrak{g}$ and $a, b \in \mathbb{C}$.  □

The above example is a special case of the following construction that arose from discussions with Vadim Schechtman.

Let $\mathfrak{G} = \mathfrak{G}_0 \oplus \mathfrak{G}_{-1} \oplus \mathfrak{G}_{-2}$ be a differential graded Lie (super-)algebra with bracket $\{\,,\,\} : \mathfrak{G}_i \times \mathfrak{G}_j \to \mathfrak{G}_{i+j}$ and with a differential $d$ of degree $+1$. Set $\tilde{\mathfrak{g}} := \mathfrak{G}_{-1}$ , $C := d(\mathfrak{G}_{-2})$, and define bilinear maps $[\,,\,] : \tilde{\mathfrak{g}} \times \tilde{\mathfrak{g}} \to \tilde{\mathfrak{g}}$ and $\langle\,,\,\rangle : \tilde{\mathfrak{g}} \times \tilde{\mathfrak{g}} \to \mathfrak{G}_{-2}$ by the formulas:

$$[x,y] = \{x, dy\} - \{y, dx\} \quad,\quad \langle x,y\rangle = \{x,y\} \tag{8.4}$$

Conversely, let $(\tilde{\mathfrak{g}}, C, [\,,\,])$ be a quasi-Lie algebra equipped with a symmetric pairing:

$$\langle\,,\,\rangle : S^2\tilde{\mathfrak{g}} \to C \tag{8.5}$$

Associate to this data the graded vector space $\mathfrak{G} := \tilde{\mathfrak{g}}/C \oplus \tilde{\mathfrak{g}} \oplus C (= \mathfrak{G}_0 \oplus \mathfrak{G}_{-1} \oplus \mathfrak{G}_{-2})$ with differential $d$ of degree $+1$ given respectively by the imbedding $C \hookrightarrow \tilde{\mathfrak{g}}$ , and by the natural projection $\tilde{\mathfrak{g}} \to \tilde{\mathfrak{g}}/C$ . Further, define a bracket $\{,\}$ by the formulas:

$$\{x,y\} = \langle x,y\rangle \quad,\quad \{x, dy\} = [x,y] + \langle x,y\rangle \quad,\quad x,y \in \tilde{\mathfrak{g}} \tag{8.6}$$

Then one has the following result which is proved by a straightforward computation.

**Lemma 8.7.** (i) *The operations (8.4) arising from a d.g. algebra $\mathfrak{G}$ make the triple $(\tilde{\mathfrak{g}}, C, [\,,\,])$ a quasi-Lie algebra with an additional symmetric pairing $\langle\,,\,\rangle$. Furthermore, the following identities hold, for any $x, y, z \in \tilde{\mathfrak{g}}$:*

$$\langle [x,y], z \rangle + \langle y, [x,z] \rangle = 0 \quad , \quad J(x,y,z) = d\langle [x,y], z \rangle \tag{8.7.1}$$

(ii) *Conversely, let $(d : C \hookrightarrow \tilde{\mathfrak{g}}, [\,,\,])$ be a quasi-Lie algebra equipped with a pairing (8.5) such that (8.7.1) holds. Then the bracket (8.6) on the associated d.g.-vector space $\mathfrak{G}$ is well-defined and puts on $\mathfrak{G}$ the unique structure of a Lie d.g.-algebra.* □

**Remark.** Lemma 8.7 is reminisent in spirit to a result of [Dr]. Namely, it shows that giving a bracket (and a bit of additional data) $\{,\} : \mathfrak{G}_0 \times \mathfrak{G}_{-1} \to \mathfrak{G}_{-1}$ satisfying the Jacobi identity but without any symmetry condition is equivalent, essentially, to giving an anti-symmetric bracket (and a bit of additional data) $[\,,\,] : \tilde{\mathfrak{g}} \times \tilde{\mathfrak{g}} \to \tilde{\mathfrak{g}}$ that does not satisfy the Jacobi identity. □

Recall now the geometric setup of the previous section. Thus, $X$ is a $d$-dimensional compact complex manifold, $G$ is a complex semisimple Lie group, $P$ is a principal holomorphic $G$-bundle on $X$, and $\mathfrak{g}_P$, resp. $\mathfrak{g}_P^*$, stands for the associated vector bundle corresponding to the adjoint, resp. coadjoint, representation of $G$. We are going to associate to $P$ a sheaf of quasi-Lie algebras on $X$.

Let $\mathcal{A}_P$ denote the *Atiyah algebra* of $P$, the sheaf (on $X$) of $G$-invariant vector fields on the total space of the $G$-bundle $P$. Any $G$-invariant vector field on $P$ gives rise, via the projection to the base, to a vector field on $X$. This way we get a natural short exact sequence of Lie algebra sheaves on $X$:

$$0 \to \mathfrak{g}_P \to \mathcal{A}_P \to \mathcal{T}_X \to 0 \tag{8.8}$$

From now on we fix a $G$-invariant (not necessarily non-degenerate) bilinear form $c \in S^2\mathfrak{g}^*$. We use the same notation, $c$, for the correspoding $G$-equivariant map $\mathfrak{g} \to \mathfrak{g}^*$ and for the induced vector bundle morphism $c : \mathfrak{g}_P \to \mathfrak{g}_P^*$. The latter morphism gives the following diagram of sheaf extensions whose first row is the pull-back of the second, and the second row is obtained by dualizing exact sequence (8.8)

$$\begin{array}{ccccccccc} 0 & \longrightarrow & \Omega^1_X & \longrightarrow & \tilde{\mathfrak{g}}_P & \longrightarrow & \mathfrak{g}_P & \longrightarrow & 0 \\ & & \| & & \downarrow & & \downarrow c & & \\ 0 & \longrightarrow & \Omega^1_X & \longrightarrow & \mathcal{A}^*_P & \longrightarrow & \mathfrak{g}^*_P & \longrightarrow & 0 \end{array}$$

Observe next that the sheaf $AutP$, the "gauge group", acts by automorphisms on all the objects in the above diagram. That action gives rise, by differentiation, to the infinitesimal gauge action of the Lie algebra sheaf $Lie(AutP) = \mathfrak{g}_P$ on all the objects involved. In particular, the $\mathfrak{g}_P$-action on $\tilde{\mathfrak{g}}_P$ arising in this way gives a $\mathbb{C}$-bilinear map $b : \mathfrak{g}_P \times \tilde{\mathfrak{g}}_P \to \tilde{\mathfrak{g}}_P$. The map $b$ is $\mathcal{O}_X$-linear with respect to the second argument and is a first order differential operator with respect to the first argument. We now define a (non $\mathcal{O}_X$-linear) skew-symmetric bracket on the sheaf $\tilde{\mathfrak{g}}_P$ by the following formula, where $\pi : \tilde{\mathfrak{g}}_P \to \mathfrak{g}_P$ denotes the projection.

$$[x, y] = \tfrac{1}{2}\big(b(\pi(x), y) - b(\pi(y), x)\big) \quad , \quad x, y \in \tilde{\mathfrak{g}}_P \tag{8.9}$$

Let $\omega(x, y, z) = c(x, [y, z])$ be the canonical $AdG$-invariant 3-form on the Lie algebra $\mathfrak{g}$ associated with $c \in S^2\mathfrak{g}$. By $G$-invariance, the 3-form on $\mathfrak{g}$ gives rise to an $\mathcal{O}_X$-linear morphism $\Lambda^3 \mathfrak{g}_P \to \mathcal{O}_X$. We often pull back the forms $c$ and $\omega$ from $\mathfrak{g}_P$ to $\tilde{\mathfrak{g}}_P$ via the projection $\pi$, thus view them as being defined on $\tilde{\mathfrak{g}}_P$.

**Lemma 8.10.** *For any $x, y, z \in \tilde{\mathfrak{g}}_P$, we have an equality (cf. Lemma 8.6) :*

$$[x, [y, z]] + [z, [x, y]] + [y, [z, x]] = dc(x, [y, z]) \in \Omega^1_X \subset \tilde{\mathfrak{g}}_P$$

**Corollary 8.11.** *The bracket $[\,,\,]$ makes the pair $\Omega^1_X \subset \tilde{\mathfrak{g}}_P$ a quasi-Lie algebra sheaf.*

*Proof of Lemma 8.10.* Observe first that the gauge action is clearly trivial on the subsheaf $\Omega^1_X \subset \tilde{\mathfrak{g}}_P$. Therefore the map $b$, hence the bracket $[\,,\,]$, descends to a (non-symmetric) pairing $b : \mathfrak{g}_P \times \mathfrak{g}_P \to \tilde{\mathfrak{g}}_P$. Furthermore, the projection $x, y \mapsto \pi(b(x, y)) \in \mathfrak{g}_P$ is given by the adjoint action of $\mathfrak{g}_P$ on itself. Hence the induced bracket $\pi \cdot [\,,\,] : \mathfrak{g}_P \times \mathfrak{g}_P \to \mathfrak{g}_P$ is nothing but the bracket arising from the natural Lie algebra structure on $\mathfrak{g}_P$.

Next, following [BS], set $\mathfrak{G}_0 = \mathfrak{g}_P$, $\mathfrak{G}_{-1} = \tilde{\mathfrak{g}}_P$ and $\mathfrak{G}_{-2} = \mathcal{O}_X$. Define the differential $d : \mathfrak{G}_{-2} \to \mathfrak{G}_{-1}$ to be the deRham differential $\mathcal{O}_X \to \Omega^1_X$ composed with the imbedding $\Omega^1_X \hookrightarrow \tilde{\mathfrak{g}}_P$, and the differential $\mathfrak{G}_{-1} \to \mathfrak{G}_0$ to be the projection $\pi$. Further, define a bracket $\{\,,\,\}$ on $\mathfrak{G} = \mathfrak{G}_0 \oplus \mathfrak{G}_{-1} \oplus \mathfrak{G}_{-2}$ as follows. Let the bracket on $\mathfrak{G}_0$ be the natural one arising from the Lie algebra structure on $\mathfrak{g}_P$. For $x \in \mathfrak{G}_0$ and $y \in \mathfrak{G}_{-1}$, set $\{x, y\} = \operatorname{ad}_x y$. For $x, y \in \mathfrak{G}_{-1}$, set $\{x, y\} = c(x, y)$, and let the $\mathfrak{G}_0$-action on $\mathfrak{G}_{-2}$ be the trivial one. It was shown in [BS] that the bracket $\{\,,\,\}$ and the differential $d$ make $\mathfrak{G}$ a differential graded Lie super-algebra. Hence, Lemma 8.6(i) applies, and Lemma 8.10 follows. $\square$

We will now give another, more transparent, interpretation of the above defined quasi-Lie algebra bracket $[\,,\,]$ on $\tilde{\mathfrak{g}}_P$ assuming that the invariant form $c \in S^2 \mathfrak{g}^*$ is non-degenerate. To that end, we begin with an alternative construction of the sheaf $\tilde{\mathfrak{g}}_P$.

Let $(\Omega^1_P)^G$ denote the sheaf on $X$ of $G$-invaiant 1-forms on the total space of the $G$-bundle $P$. The tangent space to the fibre at a point of $P$ may be identified, via the $G$-action, with Lie algebra $\mathfrak{g}$. Hence, the restriction of a 1-form on $P$ to the vectors tangent to the fibres yields a function on $\mathfrak{g} \times P$ which is linear with respect to the first factor. This way one gets a morphism of coherent $\mathcal{O}_X$-sheaves $\rho : (\Omega^1_P)^G \to (\mathfrak{g}^* \otimes \mathcal{O}_P)^G$. The kernel of $\rho$ consists of 1-forms on $P$ that are pull-backs of 1-forms on $X$. Thus, there is a canonical short exact sequence.

$$0 \to \Omega^1_X \to (\Omega^1_P)^G \xrightarrow{\rho} (\mathfrak{g}^* \otimes \mathcal{O}_P)^G \to 0 \qquad (8.12)$$

Observe that the sheaf $(\mathfrak{g}^* \otimes \mathcal{O}_P)^G$ is nothing but the sheaf of sections of the associated vector bundle $\mathfrak{g}^*_P$, so that the exact sequence above is nothing but the dual of (8.8). Further, given a section $f \in \mathfrak{g}^* \otimes \mathcal{O}_P$, let $df$ denote the exterior differential of $f$ viewed as a $\mathfrak{g}^*$-valued 1-form on $P$. Thus, we get

the following map

$$\tilde{d} : \mathfrak{g}_P^* = (\mathfrak{g}^* \otimes \mathcal{O}_P)^G \xrightarrow{d} (\mathfrak{g}^* \otimes \Omega^1{}_P)^G$$

Note next that the non-degenerate form $c$ on $\mathfrak{g}$ induces a vector bundle isomorphism $c : \mathfrak{g}_P \xrightarrow{\sim} \mathfrak{g}_P^*$. Write $\langle , \rangle$ for the pairing $\mathfrak{g}_P^* \times \mathfrak{g}_P^* \to \mathcal{O}_X$ arising from this isomorphism (i.e. the inverse of the pairing $c : \mathfrak{g}_P \times \mathfrak{g}_P \to \mathcal{O}_X$). Form the composition

$$\beta : \mathfrak{g}_P^* \otimes \mathfrak{g}_P^* \xrightarrow{\tilde{d} \otimes \mathrm{Id}} (\mathfrak{g}^* \otimes \Omega^1{}_P)^G \otimes (\mathfrak{g}^* \otimes \mathcal{O}_P)^G \xrightarrow{\langle , \rangle} \Omega^1{}_P{}^G$$

Write $x \mapsto \bar{x}$ for the projection $\rho : (\Omega_P^1)^G \to \mathfrak{g}_P^*$. We define a skew-symmetric bracket $(\Omega_P^1)^G \times (\Omega_P^1)^G \to (\Omega_P^1)^G$ by the following formula

$$x, y \mapsto [x, y] := 1/2 \bigl( \beta(\bar{x}, \bar{y}) - \beta(\bar{y}, \bar{x}) \bigr) \tag{8.13}$$

Clearly, the subsheaf $\Omega_X^1$ is contained in the center of the bracket $[\,,\,]$. The following lemma explains connections between the brackets (8.9) and (8.13); for $X = pt$ the last claim of the lemma boils down to Maurer-Cartan equations.

**Lemma 8.14.** *There is a natural isomorphism of extensions*

$$\begin{array}{ccccccccc}
0 & \longrightarrow & \Omega_X^1 & \longrightarrow & (\Omega_P^1)^G & \longrightarrow & \mathfrak{g}_P & \longrightarrow & 0 \\
& & \| & & \| & & \| c & & \\
0 & \longrightarrow & \Omega_X^1 & \longrightarrow & \tilde{\mathfrak{g}}_P & \xrightarrow{\rho} & \mathfrak{g}_P^* & \longrightarrow & 0
\end{array}$$

*Moreover, the bracket (8.13) coincides, via the above isomorphism $\tilde{\mathfrak{g}}_P \simeq (\Omega_P^1)^G$, with the quasi-Lie algebra structure (8.9) on the sheaf $\tilde{\mathfrak{g}}_P$.* $\square$

EXAMPLE 8.15. Let $P$ be the trivial $G$-bundle so that $\mathfrak{g}_P \simeq \mathfrak{g}_X := \mathfrak{g} \otimes \mathcal{O}_X$, and there is a splitting $\tilde{\mathfrak{g}}_P = \mathfrak{g}_X \oplus \Omega_X^1$. Then the bracket $[\,,\,]$ on $\tilde{\mathfrak{g}}_P$ is given by the following explicit formula:

$$[x \oplus \alpha,\, y \oplus \beta] \;=\; [x, y] \oplus c(dx, y) - c(x, dy) \qquad,\qquad x, y, [x, y] \in \mathfrak{g}_X\,,\; \alpha, \beta \in \Omega_X^1$$

where $c \in S^2 \mathfrak{g}^*$.

## 9. Differential operators on Determinant bundles

We extend here the results of [BG] to 1-st and 2-d order differential operators on a Determinant bundle on the moduli space of $G$-bundles. Generalization to higher order operators involves more sophisticated techniques of homotopy Lie algebras and will be published elsewhere. I take an opportunity to express my gratitude to Sasha Beilinson who allowed me generously to present below some of his ideas.

The notation of sections 7 and 8 are in force throughout. We assume in addition that the semisimple group $G$ is connected and simply-connected, and the manifold $X$ is 1-dimensional, more precisely, it is a smooth compact algebraic curve of genus $g > 1$. Thus we have $\Omega_X^1 = \Omega_X$.

Let $\det H$ denote the top exterior power of a finite dimensional vector space $H$. Given a holomorphic vector bundle $E$ on the curve $X$ define a 1-dimensional vector space $\mathrm{Det} E$ by the formula

$$\mathrm{Det} E = \det H^0(X, E) \otimes \det H^1(X, E)^*$$

From now on we fix a finite dimensional irreducible representation $r: G \to GL(V)$. We associate to it two different objects. The first one is the quasi-Lie algebra sheaf $\tilde{\mathfrak{g}}_P$ on $X$, cf. (8.8)-(8.9), associated to the element $c = c_r \in (S^2\mathfrak{g}^*)^G$ given by the formula

$$c : x, y \mapsto \mathrm{Tr}(r(x) \cdot r(y))$$

The second object is a holomorphic line bundle $\mathfrak{Det} = \mathfrak{Det}(V)$ on the moduli space $\mathcal{M}$ of principal holomorphic $G$-bundles on $X$. The fiber of $\mathfrak{Det}$ at a point $P \in \mathcal{M}$ is, essentially by definition, the 1-dimensional vector space $\mathrm{Det}(V_P)$, the determinant, of the associated vector bundle on $X$ corresponding to the $G$-module $V$. We will be concerned with the (locally free) sheaves, $\mathrm{Diff}_i(\mathfrak{Det})$, of holomorphic differential operators of order $\leq i$ acting on sections of the line bundle $\mathfrak{Det}$. Clearly, $\mathrm{Diff}_0(\mathfrak{Det}) = \mathcal{O}_\mathcal{M}$. For $i = 1$ geometric fibers of the sheaf $\mathrm{Diff}_1(\mathfrak{Det})$ are described by the following result, where the subscript '$P$' stands for the geometric fiber at $P$.

**Theorem 9.1 [BS].** *Let $\mathcal{M}$ be the moduli space of principal holomorphic $G$-bundles on a smooth compact algebraic variety $X$. Then, for any $G$-bundle* $P \in \mathcal{M}$ without infinitesimal automorphisms, i.e. such that $H^0(X, \mathfrak{g}_P) = 0$, there is a canonical isomorphism of vector spaces

$$\mathrm{Diff}_1(\mathfrak{Det})_{|P} = H^1(X, \tilde{\mathfrak{g}}_P) \qquad \square$$

Remark 9.2. The isomorphism of the theorem fits into the following natural commutative diagram whose second row is a fragment of the cohomology long exact sequence arising from (8.12).

$$\begin{array}{ccccccccc}
0 & \longrightarrow & \mathrm{Diff}_0(\mathfrak{Det})_{|P} & \longrightarrow & \mathrm{Diff}_1(\mathfrak{Det})_{|P} & \longrightarrow & T_P\mathcal{M} & \longrightarrow & 0 \\
& & \| & & \| & & \| & & \\
0 & \longrightarrow & H^1(X, \Omega^1_P) & \longrightarrow & H^1(X, \tilde{\mathfrak{g}}_P) & \longrightarrow & H^1(X, \mathfrak{g}_P) & \longrightarrow & 0
\end{array}$$

$\square$

It will be convenient for us to dualize the isomorphism of theorem 9.1. Using the Serre duality we obtain from the theorem the following canonical isomorphism:

$$\mathrm{Diff}_1(\mathfrak{Det})_{|P} = H^0(X, \tilde{\mathfrak{g}}_P^* \otimes \Omega^1_X)^* \qquad (9.1.1)$$

The goal of the rest of this section is to give a similar description of the fibers of $\mathrm{Diff}_2(\mathfrak{Det})$.

We recall some basic concepts concerning holomorphic differential operators. For any vector bundles $E, F$ on a smooth variety $Y$, let $\mathrm{Diff}_i(E, F)$ denote the sheaf of holomorphic differential operators from $E$ to $F$ of order $\leq i$. If $E = F$ we simply write $\mathrm{Diff}_i(F)$, for short. Differential operators are related to jet-bundles via a canonical sheaf isomorphism

$$\mathrm{Diff}_i(E, F) \simeq \mathfrak{J}^i(E)^* \otimes F \qquad, \quad i \geq 0$$

where $\mathfrak{J}^i(E)$ stands for the sheaf of $i$-jets of sections of $E$. Given a differential operator $u \in \mathrm{Diff}_i(E, F)$, one defines in a canonical way the *adjoint* operator $u^t \in \mathrm{Diff}_i(F^* \otimes \Omega_Y, E^* \otimes \Omega_Y)$. Introduce the notation $E^\vee := E^* \otimes \Omega_Y$. In the special case $F = E^\vee$ the assignment $u \mapsto u^t$ gives rise to an involution on the sheaf $\mathrm{Diff}_i(E, E^\vee)$. Let $\mathrm{Diff}_i^\pm(E, E^\vee)$ denote the $(\pm 1)$-eigenspace of that involution, the subsheaf of *selfadjoint*, resp. *anti-selfadjoint*, operators. Clearly we have $\mathrm{Diff}_0^\pm(E, E^\vee) = \mathrm{Hom}^\pm(E, E^*\otimes\Omega_Y)$ is the sheaf of selfadjoint (resp. anti-selfadjoint) $\mathcal{O}_Y$-linear operators. Thus,

$$\mathrm{Hom}^+(E, E^*\otimes\Omega_Y) = S^2 E^* \otimes \Omega_Y \quad \text{and} \quad \mathrm{Hom}^-(E, E^\vee) = \wedge^2 E^* \otimes \Omega_Y$$

Observe that, for any $i \geq 1$, we have $\mathrm{Diff}_{i-1}^-(E, E^\vee) \subset \mathrm{Diff}_i^-(E, E^\vee)$, and there is a natural isomorphism

$$\mathrm{Diff}_i^-(E, E^\vee)/\mathrm{Diff}_{i-1}^-(E, E^\vee) = \mathrm{Hom}^\pm(E, E^\vee) \otimes S^i \mathcal{T}_Y$$

where "+" sign is taken for $i$ even, "-" sign is taken for $i$ odd, $\mathcal{T}_Y$ denotes the tangent sheaf. We write $u \mapsto \hat{\sigma}(u)$ for the corresponding projection $\mathrm{Diff}_i^-(E, E^\vee) \to \mathrm{Diff}_i^-(E, E^\vee)/\mathrm{Diff}_{i-1}^-(E, E^\vee)$, the principal symbol map. Recall further the canonical isomorphism $\mathcal{T}_Y \otimes \Omega_Y \simeq \Omega_Y^{d-1}$ where $d = \dim Y$. Applying the isomorphisms above for $i = 0, 1$, one obtains the following short exact sequence

$$0 \to \wedge^2 E \otimes \Omega_Y^d \to \mathrm{Diff}_1^-(E, E^\vee) \xrightarrow{\hat{\sigma}} S^2 E \otimes \Omega_Y^{d-1} \to 0 \tag{9.3}$$

Assume now that $\dim Y = 1$, the case we are mainly interested in. Let $D \simeq Y$ be the diagonal in the Cartesian product $Y \times Y$. For any integer $k$, we use the standard notation $\mathcal{O}_{Y\times Y}(k \cdot D)$ for the sheaf (on $Y \times Y$) of regular functions on $(Y \times Y) \setminus D$ with pole of order $\leq k$ at the diagonal. Given any $\mathcal{O}_{Y\times Y}$-sheaf $\mathcal{F}$, we write $\mathcal{F}(k) := \mathcal{F} \otimes \mathcal{O}_{Y\times Y}(k \cdot D)$.

Let $E$ and $F$ be holomorphic vector bundles on $Y$ and $\phi$ a local section of the sheaf $(E^\vee \boxtimes F)(i+1)$ at a point $x \in D \subset Y \times Y$. Define a (not necessarily $\mathcal{O}_X$-linear) sheaf morphism $u_\phi : E \to F$ by the following formula in which "$res_1$" stands for the integration along a small contour around $x$ in the first factor:

$$E \ni s \mapsto u_\phi(s) = res_1 \langle s, \phi \rangle \in F$$

The following direct geometric construction of differential operators on $Y$ is due to M. Sato.

**Proposition 9.4.** (i) *For any vector bundles $E$ and $F$ on $Y$ the assignment $\phi \mapsto u_\phi$ gives a canonical isomorphism of $\mathcal{O}_Y$-bimodules:*

$$\mathrm{Diff}_i(E, F) \stackrel{v}{\simeq} \left(E^\vee \boxtimes F\right)(i+1)/\left(E^\vee \boxtimes F\right)$$

(ii) *The transposition map $u \mapsto u^t$ on differential operators corresponds, via (i), to the involution of the RHS of the isomorphism given by the transposition of the two factors $Y$.*

(iii) *In terms of isomorphism (i) the principal symbol map amounts to the natural isomorphism*

$$\left(E^\vee \boxtimes F\right)(i+1)/\left(E^\vee \boxtimes F\right)(i) \xrightarrow{\sim} E^* \otimes F \otimes \Omega_Y \otimes S^i \mathcal{T}_Y$$

□

Parts (i) and (ii) of the proposition yield the following result.

**Corollary 9.5.** (i) *The sheaf $\mathrm{Diff}_i^-(E, E^\vee)$ is isomorphic naturally to the sheaf of symmetric sections of $(E^\vee)^{\boxtimes 2}(i+1)/(E^\vee)^{\boxtimes 2}$.*

(ii) *Any vector bundle morphism $E_1 \to E_2$ induces a natural order preserving morphism $\mathrm{Diff}^-(E_2, E_2^\vee) \to \mathrm{Diff}^-(E_1, E_1^\vee)$.* □

We can now return to differential operators on the moduli space $\mathcal{M}$ of holomorphic $G$-bundles on a complex manifold $X$. Assume for the moment that $\dim X$ is arbitrary. Recall the notation of the previous section, in particular definition (8.9) of a quasi-Lie algebra structure (8.9) on $\tilde{\mathfrak{g}}_P$. For any fixed $a \in \mathfrak{g}_P$, the assignment $\mathfrak{g}_P \ni x \mapsto b(a, x) \in \tilde{\mathfrak{g}}_P$ is $\mathcal{O}_X$-linear. Therefore, for any section $\tilde{a}^\vee \in \tilde{\mathfrak{g}}_P^*$, the pairing $\mathfrak{g}_P \ni x \mapsto \langle \tilde{a}^\vee, b(da, x) \rangle$ is $\mathcal{O}_X$-linear. Hence, there is a section $\delta(\tilde{a}^\vee, a) \in \mathfrak{g}_P^*$ such that $\langle \tilde{a}^\vee, b(a, x) \rangle = \langle \delta(\tilde{a}^\vee, a), x \rangle$, for any $x$. Furthermore, for fixed $\tilde{a}^\vee$, the map $a \mapsto \delta(\tilde{a}^\vee, a)$ is given by a first order differential operator $\delta(\tilde{a}^\vee) \in \mathrm{Diff}_1(\mathfrak{g}_P, \mathfrak{g}_P^*)$. The assignment $\tilde{a}^\vee \mapsto \delta(\tilde{a}^\vee)$ arising in this way yields an $\mathcal{O}_X$-linear morphism of sheaves $\tilde{\mathfrak{g}}_P^* \to \mathrm{Diff}_1(\mathfrak{g}_P, \mathfrak{g}_P^*)$. Tensoring each side of it by $\Omega_X$ one obtains a morphism

$$\delta : \quad \tilde{\mathfrak{g}}_P^* \otimes \Omega_X \longrightarrow \mathrm{Diff}_1(\mathfrak{g}_P, \mathfrak{g}_P^* \otimes \Omega_X) \tag{9.6}$$

which is, in a sense, dual to the first half of the bracket (8.9). Define the following (non $\mathcal{O}_X$-linear) *cobracket* map:

$$\Delta = 1/2(\delta - \delta^t) : \quad \tilde{\mathfrak{g}}_P^\vee \longrightarrow \mathrm{Diff}_1^-(\mathfrak{g}_P, \mathfrak{g}_P^\vee) \tag{9.7}$$

Dualizing the basic exact sequence

$$0 \to \Omega_X^1 \to \tilde{\mathfrak{g}}_P \to \mathfrak{g}_P \to 0 \tag{9.8.1}$$

and tensoring by $\Omega_X$ one obtains, via the isomorphism above, a canonical short exact sequence

$$0 \to \mathfrak{g}_P^* \otimes \Omega_X \to \tilde{\mathfrak{g}}_P^* \otimes \Omega_X \xrightarrow{\sigma} \Omega_X^{d-1} \qquad d = \dim X \tag{9.8.2}$$

Recall that the composition of the bracket (8.9) with the projection $\pi : \tilde{\mathfrak{g}}_P \twoheadrightarrow \mathfrak{g}_P$ is the standard $\mathcal{O}_X$-bilinear bracket on the Lie algebra $\mathfrak{g}_P$. Hence the cobracket (9.7) being restricted to the subsheaf $\mathfrak{g}_P^* \otimes \Omega_X \subset \tilde{\mathfrak{g}}_P^* \otimes \Omega_X$ becomes the morphism

$$\Delta: \ \mathfrak{g}_P^* \otimes \Omega_X \to (\mathfrak{g}_P^* \wedge \mathfrak{g}_P^*) \otimes \Omega_X = \mathrm{Diff}_0^-(\mathfrak{g}_P, \mathfrak{g}_P^* \otimes \Omega_X)$$

induced by the ordinary co-bracket map $\mathfrak{g}_P^* \to \mathfrak{g}_P^* \wedge \mathfrak{g}_P^*$, twisted by $\Omega_X$.

Further, let $c : \Omega_X^{d-1} \to (\mathfrak{g}_P^* \otimes \mathfrak{g}_P^*) \otimes \Omega_X^{d-1}$ be the morphism induced by the map $\mathcal{O}_X \to \mathfrak{g}_P^* \otimes \mathfrak{g}_P^*$ sending the function 1 to the section corresponding to the invariant form $c \in S^2\mathfrak{g}^* \subset \mathfrak{g}^* \otimes \mathfrak{g}^*$.

**Lemma 9.9.** *The following natural diagram commutes:*

$$\begin{array}{ccccccccc}
0 & \longrightarrow & \mathfrak{g}_P^\vee & \longrightarrow & \tilde{\mathfrak{g}}_P^\vee & \xrightarrow{\sigma} & \Omega_X^{d-1} & \longrightarrow & 0 \\
& & \downarrow{\scriptstyle \Delta} & & \downarrow{\scriptstyle \Delta} & & \downarrow{\scriptstyle c} & & \\
0 & \longrightarrow & \mathrm{Diff}_0^-(\mathfrak{g}_P, \mathfrak{g}_P^\vee) & \longrightarrow & \mathrm{Diff}_1^-(\mathfrak{g}_P, \mathfrak{g}_P^\vee) & \xrightarrow{\hat{\sigma}} & S^2\mathfrak{g}_P^* \otimes \Omega_X^{d-1} & \longrightarrow & 0
\end{array}$$

□

The above can be made more explicit in the special case $d = 1$. Proposition 9.4 and corollary 9.5 yield the following strengthenning of lemma 9.7.

**Lemma 9.10.** *For a $G$-bundle $P$ on a one-dimensional manifold $X$ there is a natural commutative diagram:*

$$
\begin{array}{ccccccccc}
0 & \to & \mathfrak{g}_P^\vee & \to & \tilde{\mathfrak{g}}_P^\vee & \stackrel{\sigma}{\to} & \mathcal{O}_X & \to & 0 \\
& & \downarrow \Delta & & \downarrow \Delta & & \downarrow c & & \\
0 & \to & \mathrm{Diff}_0^-(\mathfrak{g}_P, \mathfrak{g}_P^\vee) & \to & \mathrm{Diff}_1^-(\mathfrak{g}_P, \mathfrak{g}_P^\vee) & \stackrel{\hat{\sigma}}{\to} & S^2 \mathfrak{g}_P^* & \to & 0 \\
& & \| (9.3) & & \| 9.4(i) & & \| 9.4(iii) & & \\
0 & \to & \wedge^2 \mathfrak{g}_P^\vee \otimes \Omega_X^1 & \to & (\wedge^2 \mathfrak{g}_P^\vee)(2D) / (\wedge^2 \mathfrak{g}_P^\vee) & \to & S^2 \mathfrak{g}_P^* & \to & 0
\end{array}
$$

□

The rest of the section is devoted to the formulation of main result. We keep the assumption that $X$ is a smooth compact algebraic curve and observe that there is a natural isomorphism

$$\Omega_X^1 \boxtimes \Omega_X^1 \simeq sign \otimes \Omega_{X^2}^2,$$

The "*sign*" symbol on the rihgt is inserted to make the isomorphism commute with the natural $\Sigma_2$-actions on both sides (the action on the left is given by permutation, while the action on the right is induced by the transposition of the Cartesian square $X \times X$. We introduce further the sheaf $(\tilde{\mathfrak{g}}_P^\vee)^{\boxtimes 2}$ on $X \times X$ which may be regarded, in view of the above isomorphism, as a sheaf of $(\tilde{\mathfrak{g}}_P^*)^{\boxtimes 2}$-valued 2-forms. Therefore, the sheaf $(\tilde{\mathfrak{g}}_P^\vee)^{\boxtimes 2}(2D)$ of sections with second order pole at the diagonal contains $(\tilde{\mathfrak{g}}_P^*)^{\boxtimes 2} \otimes \Omega_{X^2, X^2 \setminus D}$, the sheaf of $(\tilde{\mathfrak{g}}_P^*)^{\boxtimes 2}$-valued 2-forms on $X \times X$ with logarithmic singularities at $D$. Observe further that $Sym^2 X := X^2 / \Sigma_2$, the symmetric square of $X$, is a smooth complex surface, and $\Sigma_2$-symmetric sections of $(\tilde{\mathfrak{g}}_P^\vee)^{\boxtimes 2}(2D)$ form a well-defined sheaf, $Sym^2 \tilde{\mathfrak{g}}_P^\vee(2D)$, on $Sym^2 X$. Usually, we identify a local section of this sheaf with the corresponding section of $(\tilde{\mathfrak{g}}_P^\vee)^{\boxtimes 2}(2D)$ on $X^2$ itself.

We are going to define some important morphisms. First, we have a natural projection $sing : (\tilde{\mathfrak{g}}_P^\vee)^{\boxtimes 2}(2D) \to (\tilde{\mathfrak{g}}_P^\vee)^{\boxtimes 2}(2D)/(\tilde{\mathfrak{g}}_P^\vee)^{\boxtimes 2}$ taking the "singular part" at the diagonal. Composing it with the isomorphism $v$ of proposition 9.4(i) applied to $E = \tilde{\mathfrak{g}}_P$ and $F = \tilde{\mathfrak{g}}_P^\vee$, and of the isomorphism gives rise to a morphism

$$u : (\tilde{\mathfrak{g}}_P^\vee)^{\boxtimes 2}(2D) \xrightarrow{sing} (\tilde{\mathfrak{g}}_P^\vee)^{\boxtimes 2}(2D)/(\tilde{\mathfrak{g}}_P^\vee)^{\boxtimes 2} \xrightarrow[\simeq]{v} \mathrm{Diff}_1^-(\tilde{\mathfrak{g}}_P, \tilde{\mathfrak{g}}_P^\vee)$$

Being restricted to symmetric sections, it yields the following commutative triangle

$$\begin{array}{c} Sym^2 \tilde{\mathfrak{g}}_P^\vee(2D) \xrightarrow{sing} Sym^2 \tilde{\mathfrak{g}}_P^\vee(2D)/Sym^2 \tilde{\mathfrak{g}}_P^\vee \\ \searrow u \quad \quad \parallel v \\ \mathrm{Diff}_1^-(\tilde{\mathfrak{g}}_P, \tilde{\mathfrak{g}}_P^\vee) \end{array} \quad (9.11)$$

Second, let $\tilde{\mathfrak{g}}_P^\circ$ be the inverse image of the constant sheaf $\mathbb{C}_X \subset \mathcal{O}_X$ under the natural projection $\sigma : \tilde{\mathfrak{g}}_P^* \otimes \Omega_X^1 \to \mathcal{O}_X$, see top row of diagram 9.10. We have a short exact sequence

$$0 \to \tilde{\mathfrak{g}}_P^\vee \to \tilde{\mathfrak{g}}_P^\circ \xrightarrow{\sigma} \mathbb{C}_X \to 0$$

For any $\omega \in \tilde{\mathfrak{g}}_P^\circ$ the first order differential operator $\Delta(\omega)$, see lemma 9.10, has constant principal symbol. Note also that replacing $\tilde{\mathfrak{g}}_P^\vee$ by $\tilde{\mathfrak{g}}_P^\circ$ doesn't affect the global cohomology group : $H^0(X, \tilde{\mathfrak{g}}_P^\vee) = H^0(X, \tilde{\mathfrak{g}}_P^\circ)$, due to the vanishing condition $H^0(X, \mathfrak{g}_P) = 0$ and the equality $H^0(X, \mathcal{O}_X) = \mathbb{C}$.

Note finally, that the projection $\sigma$ induces a morphism

$$\sigma \boxtimes Id : Sym^2 \tilde{\mathfrak{g}}_P^\vee(2D) \hookrightarrow (\tilde{\mathfrak{g}}_P^\vee)^{\boxtimes 2}(2D) \to \mathcal{O}_X \boxtimes \tilde{\mathfrak{g}}_P^\vee \quad (9.12)$$

Motivated by the constructions of §§7 − 8 we now introduce the following

**Definition 9.13.** Notation: for any $\omega \in (\tilde{\mathfrak{g}}_P^\vee)^{\boxtimes 2}(2D)$ such that $(\sigma \boxtimes Id)(\omega) \in 1 \boxtimes \tilde{\mathfrak{g}}_P^\vee$ write $(\sigma \boxtimes Id)(\omega) = 1 \boxtimes \alpha(\omega)$ where $\alpha(\omega)$ is a uniquely determined section of $\tilde{\mathfrak{g}}_P^\vee$.

Define a sheaf $\mathcal{G}^\spadesuit$ on $Sym^2 X$ to consist of all $\omega \in \mathcal{G}^\spadesuit \in Sym^2 \tilde{\mathfrak{g}}_P^\vee(2D)$ satisfying the following two conditions:

(i) $$(\sigma \boxtimes Id)(\omega) = 1 \boxtimes \alpha(\omega) \quad \text{where} \quad \alpha(\omega) \in \tilde{\mathfrak{g}}_P^\circ \subset \tilde{\mathfrak{g}}_P^\vee$$

(ii) $u(\omega)$ and $\Delta(\alpha(\omega))$, the image of $\omega$ under the map (9.11) and the image of $\alpha(\omega)$ under map (9.7), respectively are both sections of the subsheaf $\mathrm{Diff}_1^-(\mathfrak{g}_P, \mathfrak{g}_P^\vee) \subset \mathrm{Diff}_1^-(\tilde{\mathfrak{g}}_P, \tilde{\mathfrak{g}}_P^\vee)$, imbedded via 9.5(ii); moreover, the following equality holds
$$u(\omega) = \Delta(\alpha(\omega)) \qquad \text{in} \quad \mathrm{Diff}_1^-(\mathfrak{g}_P, \mathfrak{g}_P^\vee). \quad \square$$

Here is the main result of this section.

**Theorem 9.14.** *Let $\mathcal{M}$ be the moduli space of holomorphic $G$-bundles on a connected complex compact manifold $X$. Then, for any bundle $P \in \mathcal{M}$ without infinitesimal automorphisms there is a canonical isomorphism of vector spaces*
$$\mathrm{Diff}_2(\mathfrak{Det})_{|P}^* \simeq H^0\big(Sym^2 X, \mathcal{G}^\spadesuit\big) \qquad \square$$

We will now study the structure of the sheaf $\mathcal{G}^\spadesuit$ in more detail. Note that the assignment $\omega \mapsto \alpha(\omega)$ gives a sheaf morphism $\alpha : \mathcal{G}^\spadesuit \to \tilde{\mathfrak{g}}_P^\circ$. This morphism is surjective. Futher, if $\alpha(\omega) = 0$ then $u(\omega) = \Delta(\alpha(\omega)) = 0$, by equation 9.13(ii). It follows that the kernel of the morphism $\omega \mapsto \alpha(\omega)$ is formed by sections of $(\tilde{\mathfrak{g}}_P^\vee)^{\boxtimes 2}(2D)$ that have no singularity at the diagonal. Moreover, any such section $\omega$ is $\Sigma_2$-symmetric and one has $(\sigma \boxtimes Id)(\omega) = 1 \boxtimes \alpha(\omega) = 0$. Hence, also $(Id \boxtimes \sigma)(\omega) = 0$, so that actually we find that $\omega \in Sym^2 \mathfrak{g}_P^\vee$. Thus, one gets a short exact sequence

$$0 \to Sym^2 \mathfrak{g}_P^\vee \to \mathcal{G}^\spadesuit \xrightarrow{\alpha} \tilde{\mathfrak{g}}_P^\circ \to 0 \qquad (9.15)$$

**Remark 9.16.** The natural short exact sequence
$$\mathrm{Diff}_1(\mathfrak{Det})_{|P} \hookrightarrow \mathrm{Diff}_2(\mathfrak{Det})_{|P} \twoheadrightarrow \big(\mathrm{Diff}_2(\mathfrak{Det})/\mathrm{Diff}_1(\mathfrak{Det})\big)_{|P}$$

corresponds, via theorems 9.1 and 9.14, to the dual of a fragment of the cohomology exact sequence
$$0 \to S^2 H^0(X, \mathfrak{g}_P^\vee) \to H^0(Sym^2 X, \mathcal{G}^\spadesuit) \to H^0(X, \tilde{\mathfrak{g}}_P^\circ) = H^0(X, \tilde{\mathfrak{g}}_P^\vee) \to 0$$

arising from (9.14).  □

Next we introduce the subsheaf

$$\mathcal{G}^{\clubsuit} := \{\omega \in Sym^2\mathfrak{g}_P^\vee(2D) \mid u(\omega) \in \Delta(\tilde{\mathfrak{g}}_P^\circ) \subset \mathrm{Diff}_1^-(\mathfrak{g}_P, \mathfrak{g}_P^\vee)\} \qquad (9.17)$$

The projection *sing* yields the following natural commutative diagram arising from lemma 9.10.

$$\begin{array}{ccccccccc}
0 & \longrightarrow & Sym^2\mathfrak{g}_P^\vee & \longrightarrow & \mathcal{G}^{\clubsuit} & \xrightarrow{sing} & \tilde{\mathfrak{g}}_P^\circ & \longrightarrow & 0 \\
& & \| & & \downarrow & & \downarrow \Delta & & \\
0 & \longrightarrow & Sym^2\mathfrak{g}_P^\vee & \longrightarrow & Sym^2\mathfrak{g}_P^\vee(2D) & \xrightarrow{u} & \mathrm{Diff}_1^-(\mathfrak{g}_P, \mathfrak{g}_P^\vee) & \longrightarrow & 0
\end{array} \qquad (9.18)$$

Furthermore, the top row of the diagram above is isomorphic to the pull-back of the second row via the map $\Delta$.

We want to analyze the relative position of the following subsheaves

$$Sym^2\tilde{\mathfrak{g}}_P^\circ \ , \ Sym^2\mathfrak{g}_P^\vee(2D) \ , \ \mathcal{G}^{\spadesuit} \quad \mathcal{G}^{\clubsuit} \ \subset \ Sym^2\tilde{\mathfrak{g}}_P^\vee(2D)$$

with the help of the commutative diagram below induced by (9.15) and by the inclusion $i : \mathfrak{g}_P^\vee \hookrightarrow \tilde{\mathfrak{g}}_P^\vee$

$$\begin{array}{ccccccccc}
0 & \longrightarrow & Sym^2\mathfrak{g}_P^\vee & \longrightarrow & Sym^2\mathcal{G}^{\spadesuit} & \xrightarrow{\alpha} & \tilde{\mathfrak{g}}_P^\circ & \longrightarrow & 0 \\
& & \downarrow & & \downarrow & & \downarrow \Delta & & \\
0 & \longrightarrow & Sym^2\tilde{\mathfrak{g}}_P^\vee & \longrightarrow & Sym^2\tilde{\mathfrak{g}}_P^\vee(2D) & \xrightarrow{sing} & \frac{Sym^2\tilde{\mathfrak{g}}_P^\vee(2D)}{Sym^2\tilde{\mathfrak{g}}_P^\vee} & \longrightarrow & 0 \\
& & \uparrow i & & \uparrow i & & \uparrow i & & \\
0 & \longrightarrow & Sym^2\mathfrak{g}_P^\vee & \longrightarrow & Sym^2\mathfrak{g}_P^\vee(2D) & \xrightarrow{sing} & \frac{Sym^2\mathfrak{g}_P^\vee(2D)}{Sym^2\mathfrak{g}_P^\vee} & \longrightarrow & 0
\end{array}$$

This diagram, together with (9.15) and (9.18), shows that

$$sing(\mathcal{G}^{\clubsuit}) \ = \ sing(\mathcal{G}^{\spadesuit}) \ \simeq \ \tilde{\mathfrak{g}}_P^\circ$$

Observe next that for any $\omega \in Sym^2\tilde{\mathfrak{g}}_P^\circ$ we have $(\sigma \boxtimes Id)(\omega) = 1 \boxtimes \alpha(\omega)$ where $\alpha(\omega) \in \tilde{\mathfrak{g}}_P^\circ$, cf. (9.13). Hence, the assignment $\omega \mapsto \alpha(\omega)$ gives rise to a natural extension

$$0 \to Sym^2\mathfrak{g}_P^\vee \to Sym^2\tilde{\mathfrak{g}}_P^\circ \xrightarrow{\alpha} \tilde{\mathfrak{g}}_P^\circ \to 0 \tag{9.19}$$

From the observations above one derives the following result.

**Lemma 9.20.** (i) *We have* : $\mathcal{G}^\clubsuit \cap Sym^2\tilde{\mathfrak{g}}_P^\circ = Sym^2\mathfrak{g}_P^\vee$. *Futhermore*

(ii) $\quad \mathcal{G}^\spadesuit = \{\omega = \omega_1 + \omega_2 \mid \omega_1 \in \mathcal{G}^\clubsuit,\ \omega_2 \in Sym^2\tilde{\mathfrak{g}}_P^\circ,\ sing(\omega_1) = \alpha(\omega_2)\}$

$\square$

Write

$$\mathfrak{g} \oplus_{\tilde{\mathfrak{g}}_P^\circ} Sym^2\tilde{\mathfrak{g}}_P^\circ = \{\omega_1 \oplus \omega_2 \in \mathcal{G}^\clubsuit \oplus Sym^2\tilde{\mathfrak{g}}_P^\circ \mid sing(\omega_1) = \alpha(\omega_2)\}$$

Combining both parts of the lemma we get a natural isomorphism

$$\mathcal{G}^\spadesuit \simeq (\mathcal{G}^\clubsuit \oplus_{\tilde{\mathfrak{g}}_P^\circ} Sym^2\tilde{\mathfrak{g}}_P^\circ)/Sym^2\mathfrak{g}_P^\vee \tag{9.21}$$

where $Sym^2\mathfrak{g}_P^\vee \hookrightarrow \mathcal{G}^\clubsuit \oplus Sym^2\tilde{\mathfrak{g}}_P^\circ$ is the anti-diagonal imbedding.

## 10. Flat projective connection on conformal blocks

Throughout this section we fix a positive integer $g > 1$, a connected simply-connected complex semisimple group $G$, and a finite dimensional rational representation $r : G \to GL(V)$ that does not contain the trivial representation of $G$ as a direct summand. Let $\mathcal{M}_g$ be the moduli space of smooth irreducible curves of genus $g$, as in §3. Given a curve $X \in \mathcal{M}_g$, let $\mathcal{M}_X$ be the moduli space of principal holomorphic $G$-bundles on $X$ and let $\mathfrak{Det}_{X,V} = \mathfrak{Det}(V)$ be the determinant bundle on $\mathcal{M}_X$ attached to the $G$-module $V$ as explained at the beginning of the previous section.

For any $X \in \mathcal{M}_g$, the space $H^0(\mathcal{M}_X, \mathfrak{Det}_{X,V})$ is known to be a finite dimensional vector space whose dimension is independent of $X$. Moreover, there is a holomorphic vector bundle, $\mathfrak{B}$, on $\mathcal{M}_g$, called "*the vector bundle of conformal blocks*", cf. [AW], whose fiber at a point $X \in \mathcal{M}_g$ is the vector space $H^0(\mathcal{M}_X, \mathfrak{Det}_{X,V})$. The famous result of [AW], [BK], [Fa], and [Hi] says that there is a natural holomorphic flat projective connection $\nabla$ on $\mathfrak{B}$. (A *flat projective connection* on a vector bundle $E$ is, by definition, a flat connection on the principal $PGL$-bundle corresponding to $E$). The purpose of this section is to present a new *local algebraic* construction of the connection $\nabla$ (cf. [Hi] for a local but not algebraic construction).

To begin with, we give a slightly different interpretation of the vector bundle of conformal blocks. Let $\mathbf{M}_g$ be the moduli space of all pairs $(X, P)$ where $X \in \mathcal{M}_g$ is a curve and $P$ is a stable holomorphic $G$-bundle on $X$. The first projection $(X, P) \mapsto X$ gives a fibration $p : \mathbf{M}_g \to \mathcal{M}_g$. For any $X \in \mathcal{M}_g$, the fiber $p^{-1}(X)$ is the moduli space, $\mathcal{M}_X$, of stable $G$-bundles on $X$. Further, there is a holomorphic line bundle $\mathfrak{Det}$ on the whole of $\mathbf{M}_g$ whose restriction to any fiber $p^{-1}(X) = \mathcal{M}_X$ is the determinant bundle $\mathfrak{Det}_{X,V}$ studied in the previous section. It is clear that the direct image sheaf, $p_*\mathfrak{Det}$, on $\mathcal{M}_g$ is nothing but the sheaf of sections of the vector bundle $\mathfrak{B}$.

Recall now that to the $G$-module $V$, we have associated at the beginning of the previous section, an invariant form $c \in (S^2\mathfrak{g}^*)^G$. Note that $c$ is non-degenerate, for $V$ does not contain the trivial representation as a component. Hence the form $c$ induces an invariant form $\langle , \rangle$ on $\mathfrak{g}^*$, the dual of the Lie algebra $\mathfrak{g}$. Fix a curve $X$ and a $G$-bundle $P$ on it. The invariant form $\langle , \rangle$ gives a sheaf morphism $S^2\mathfrak{g}^*_P \to \mathcal{O}_X$. Tensoring with $\Omega_X^{\otimes 2}$ yields a morphism

$$\langle , \rangle_P : S^2\mathfrak{g}^\vee_P = S^2\mathfrak{g}^*_P \otimes \Omega_X^{\otimes 2} \to \Omega_X^{\otimes 2}$$

Recall next the sheaf $\mathcal{G}^\spadesuit$ defined in 9.13. We consider the following diagram of extensions obtained by, first, restricting the sections of extension (9.15) to the diagonal $D \subset X \times X$, and then "pushing out" the resulting extension via the map $\langle , \rangle_P$ above

$$
\begin{array}{ccccccccc}
0 & \longrightarrow & Sym^2 \mathfrak{g}_P^\vee & \longrightarrow & \mathcal{G}^\spadesuit & \xrightarrow{\alpha} & \tilde{\mathfrak{g}}_P^\circ & \longrightarrow & 0 \\
 & & \downarrow{\scriptstyle restriction} & & \downarrow{\scriptstyle restriction} & & \parallel & & \\
0 & \longrightarrow & S^2 \mathfrak{g}_P^\vee & \longrightarrow & \mathcal{G}^\spadesuit|_D & \longrightarrow & \tilde{\mathfrak{g}}_P^\circ & \longrightarrow & 0 \\
 & & \downarrow{\scriptstyle \langle,\rangle_P} & & \downarrow & & \parallel & & \\
0 & \longrightarrow & \Omega_X^{\otimes 2} & \longrightarrow & \mathcal{G}_{X,P} & \longrightarrow & \tilde{\mathfrak{g}}_P^\circ & \longrightarrow & 0
\end{array}
\quad (10.1)
$$

Thus, we have defined a sheaf $\mathcal{G}_{X,P}$ on $X$ to be the extension in the last row of the diagram.

There is an analogue of formula (9.1.1) for the universal moduli space $\mathbf{M}_g$. Observe that the sheaf $\mathrm{Diff}_1(\mathfrak{Det})$ of the first order differential operators on the vector bundle $\mathfrak{Det}$ contains the subsheaf $\mathrm{Diff}_1^\uparrow(\mathfrak{Det})$ of the "vertical differential operators", acting along the fibers of the projection $p : \mathbf{M}_g \to \mathcal{M}_g$. Observe that $\mathrm{Diff}_0(\mathfrak{Det}) \subset \mathrm{Diff}_1^\uparrow(\mathfrak{Det})$. Furthermore, composing the principal symbol map $\mathrm{Diff}_1(\mathfrak{Det}) \to \mathrm{Diff}_1(\mathfrak{Det})/\mathrm{Diff}_0(\mathfrak{Det}) \simeq \mathcal{T}_{\mathbf{M}_g}$ with the natural projection $p_* : \mathcal{T}_{\mathbf{M}_g} \to \mathcal{T}_{\mathcal{M}_g}$ yields a canonical short exact sequence

$$\mathrm{Diff}_1^\uparrow(\mathfrak{Det}) \hookrightarrow \mathrm{Diff}_1(\mathfrak{Det}) \xrightarrow{\pi} p^* \mathcal{T}_{\mathcal{M}_g} \quad (10.2)$$

The result below was proved in [BS] in the special case $G = SL_n(\mathbb{C})$. It describes the locally free sheaf of the first order differential operators on the vector bundle $\mathfrak{Det}$.

Theorem 10.3 *For any curve $X \in \mathcal{M}_g$ and any $G$-bundle $P$ on $X$, there is a canonical isomorphism of vector spaces*

$$\mathrm{Diff}_1(\mathfrak{Det})_{|X,P} \simeq H^0(X, \mathcal{G}_{X,P})^* \quad , \quad (X,P) \in \mathbf{M}_g$$

*such that the short exact sequence (10.2) gets identified, via the above isomorphism and theorems 3.2 and 9.14, with the dual of the cohomology exact sequence*

$$H^0(X, \Omega_X^{\otimes 2}) \hookrightarrow H^0(X, \mathcal{G}_{X,P}) \twoheadrightarrow H^0(X, \tilde{\mathfrak{g}}_P^\circ)$$

*induced by (10.1).* □

Observe next that we have $\text{Diff}_1(\mathfrak{Det}) \cap \text{Diff}_2^\uparrow(\mathfrak{Det}) = \text{Diff}_1^\uparrow(\mathfrak{Det})$, and that the geometric fibers of the sheaf on the right are described by theorem 9.1. Further, composing the natural maps in the middle vertical column of diagram (10.1) induces a morphism of cohomology:

$$\nu : H^0(Sym^2 X, \mathcal{G}^\spadesuit) \to H^0(X, \mathcal{G}_{X,P}) \quad , \quad (X,P) \in \mathbf{M}_g \qquad (10.4)$$

From theorem 10.3 we deduce the following result that plays a crucial role in our construction of a natural flat projective connection on the vector bundle $\mathfrak{B} = p_*\mathfrak{Det}$.

**Theorem 10.5.** *For any rational $G$-module $V$, the duals to the maps (10.4) give rise, via the isomorphism of theorem 10.3, a canonical $\mathcal{O}_{\mathbf{M}_g}$-linear Lie algebra homomorphism of sheaves on $\mathbf{M}_g$*

$$\nu^* : \text{Diff}_1(\mathfrak{Det}) \to \text{Diff}_2^\uparrow(\mathfrak{Det})$$

*which restricts to the identity map on the subsheaf* $\text{Diff}_1^\uparrow(\mathfrak{Det}) = \text{Diff}_1(\mathfrak{Det}) \cap \text{Diff}_2^\uparrow(\mathfrak{Det})$. □

**Construction of the flat projective connection on $\mathfrak{B}$:** Recall the projection $p : \mathbf{M}_g \to \mathcal{M}_g$. Let $X$ be a curve and $\xi$ a germ at $X \in \mathcal{M}_g$ of a holomorphic vector field on $\mathcal{M}_g$. Choose a point $(X,P) \in \mathbf{M}_g$ over $X$ and a germ of section $\tilde{\xi} \in \text{Diff}_1(\mathfrak{Det})$ at $(X,P) \in \mathbf{M}_g$ such that $\pi(\tilde{\xi}) = \xi$ where $\pi$ denotes the map introduced in (10.2). The section $\tilde{\xi}$ is determined up to addition of a "vertical" summand $\tilde{\xi}^\uparrow \in \text{Diff}_1^\uparrow(\mathfrak{Det})$. Hence, the difference $\tilde{\xi} - \nu^*(\tilde{\xi})$ does not depend on the choice of such a lifting, due to theorem 10.5. Thus, the assigment $\xi \mapsto \tilde{\xi} - \nu * (\tilde{\xi})$ gives a well defined morphism, $\nabla$, from germs at $X \in \mathcal{M}_g$ of vector fields on $\mathcal{M}_g$ to germs at $(X,P)$ of sections of the sheaf $\text{Diff}_1(\mathfrak{Det}) + \text{Diff}_2^\uparrow$. Furthermore, the assignment $\nabla$ depends holomorphically on the point $P$ in the fiber, $p^{-1}(X)$, over $X$, so that one actually gets a map (of stalks at any $X$) of the sheaves

$$\mathcal{T}_{\mathcal{M}_g} \to p_*\Big(\text{Diff}_1(\mathfrak{Det}) + \text{Diff}_2^\uparrow\Big)$$

But such a map clearly gives rise to a flat projective connection on the vector bundle $p_*\mathfrak{Det}$. □